\documentclass[letterpaper,twocolumn,10pt]{article}
\usepackage{usenix}

\usepackage{tikz}
\usepackage{amsmath}

\usepackage{filecontents}

\usepackage{svg}
\usepackage{mdframed}
\usepackage{graphicx}
\usepackage{algorithm}
\usepackage{algorithmicx}
\usepackage{gensymb}
\usepackage[utf8]{inputenc}
\DeclareUnicodeCharacter{2212}{-}
\usepackage{mathtools}
\usepackage{algpseudocode}
\usepackage{multirow}
\usepackage{amssymb}
\usepackage{url}
\usepackage{hyperref}
\usepackage{xcolor,colortbl}
\usepackage{caption}    
\usepackage{array}
\usepackage{booktabs}
\usepackage{pifont}
\usepackage{verbatim}
\usepackage{tikz}

\usepackage{makecell}

\usepackage{amsmath, amssymb, amsfonts}
\usepackage{enumitem}
\usepackage{siunitx}
\usepackage{subcaption}

\usepackage{listings}
\usepackage{xcolor}
\definecolor{green}{rgb}{0.01, 0.75, 0.24}
\definecolor{red}{rgb}{0.76, 0.23, 0.13}
\usepackage{pifont}
\newcommand{\cmark}{\textcolor{green}{\ding{51}}}%
\newcommand{\xmark}{\textcolor{red}{\ding{55}}}%

\definecolor{red}{HTML}{FF6666}
\definecolor{green}{HTML}{67AB9F}
\definecolor{purple}{HTML}{CCCCFF}

\lstdefinestyle{highlight}{
    backgroundcolor=\color{yellow},
}

\usepackage{pifont}

\renewcommand{\paragraph}[1]{\medskip \noindent {\bf #1.}}

\newcommand{\lstbg}[3][0pt]{{\fboxsep#1\colorbox{#2}{\strut #3}}}
\lstdefinelanguage{diff}{
  basicstyle=\ttfamily\small,
  morecomment=[f][\lstbg{red!20}]-,
  morecomment=[f][\lstbg{green!20}]+,
  morecomment=[f][\textit]{@@},
}

\definecolor{mGreen}{rgb}{0,0.58,0}
\definecolor{mGray}{rgb}{0.5,0.5,0.5}
\definecolor{mPurple}{rgb}{0.58,0,0.82}
\definecolor{backgroundColour}{rgb}{0.80,0.80,0.80}

\lstdefinelanguage{markdown}{
  morekeywords={},
  sensitive=false, 
  basicstyle=\footnotesize\ttfamily,
  columns=fullflexible,
  breaklines=true,      
  frame=single, 
  morecomment=[l]{\#},
  morestring=[b]",
}

\lstdefinestyle{CStyle2}{
  backgroundcolor=\color{white},
  basicstyle=\footnotesize\ttfamily,
  columns=fullflexible,
  breakatwhitespace=false,      
  breaklines=true,                
  captionpos=b,                    
  commentstyle=\color{mGreen}, 
  extendedchars=true,              
  frame=single,                   
  keepspaces=true,             
  keywordstyle=\color{blue},      
  language=c++,                 
  numbers=left,                
  numbersep=5pt,                   
  numberstyle=\tiny\color{mGray}, 
  rulecolor=\color{mGray},        
  showspaces=false,               
  showtabs=false,                 
  stepnumber=1,                  
  stringstyle=\color{magenta},    
  tabsize=1,                      
  title=\lstname,
  morecomment=[f][\lstbg{red!20}]-,
  morecomment=[f][\lstbg{green!20}]+,
  morecomment=[f][\textit]{@@},
}

\definecolor{mGreen}{rgb}{0,0.58,0}
\definecolor{mGray}{rgb}{0.5,0.5,0.5}
\definecolor{mPurple}{rgb}{0.58,0,0.82}
\definecolor{backgroundColour}{rgb}{0.80,0.80,0.80}

\begin{document}
\lstset{
    language=[x86masm]Assembler,
    basicstyle=\ttfamily\footnotesize,
    keywordstyle=\bfseries,
    numbers=left,
    numberstyle=\tiny,
    stepnumber=1,
    breaklines=true,
    captionpos=b,
    frame=single,
    tabsize=4,
    keywords=[2]{fcomi,endrep,STMXCSR, VPBLENDMB,PSUBQ,VMOVSD,FCOMIP,VCMPPD},
}

\newcommand{\Mi}{$\mu$Arch }
\date{}

\title{\Large \bf {$\mu$}RL: Discovering Transient Execution Vulnerabilities\\ Using Reinforcement Learning}

\author{
{\rm M. Caner Tol}\\
Worcester Polytechnic Institute\\
mtol@wpi.edu
\and
{\rm Kemal Derya}\\
Worcester Polytechnic Institute\\
kderya@wpi.edu
\and
{\rm Berk Sunar}\\
Worcester Polytechnic Institute\\
sunar@wpi.edu
} 

\maketitle


\begin{abstract}
We propose using reinforcement learning to address the challenges of discovering microarchitectural vulnerabilities, such as Spectre and Meltdown, which exploit subtle interactions in modern processors. Traditional methods like random fuzzing fail to efficiently explore the vast instruction space and often miss vulnerabilities that manifest under specific conditions. To overcome this, we introduce an intelligent, feedback-driven approach using RL. Our RL agents interact with the processor, learning from real-time feedback to prioritize instruction sequences more likely to reveal vulnerabilities, significantly improving the efficiency of the discovery process.
    
We also demonstrate that RL systems adapt effectively to various microarchitectures, providing a scalable solution across processor generations. By automating the exploration process, we reduce the need for human intervention, enabling continuous learning that uncovers hidden vulnerabilities. Additionally, our approach detects subtle signals, such as timing anomalies or unusual cache behavior, that may indicate microarchitectural weaknesses. This proposal advances hardware security testing by introducing a more efficient, adaptive, and systematic framework for protecting modern processors.

When unleashed on Intel Skylake-X and Raptor Lake microarchitectures, our RL agent was indeed able to generate instruction sequences that cause significant observable byte leakages through transient execution without generating any $\mu$code assists, faults or interrupts. The newly identified leaky sequences stem from a variety of Intel instructions, e.g. including SERIALIZE, VERR/VERW, CLMUL, MMX-x87 transitions, LSL+RDSCP and LAR. These initial results give credence to the proposed approach. 

\end{abstract}

\section{Introduction}
In the past two decades, our computing systems have evolved and grown at an astounding rate. A side effect of this growth has been increased resource sharing and, with it, erosion of isolation boundaries. \textit{Multitenancy} has already been shown to be a significant security and privacy threat in shared cloud instances. VM boundaries can be invalidated either by software or hardware bugs~\cite{van2020lvi,islam2019spoiler,moghimi2020medusa,correia2018copycat} or by exploiting subtle information leakages at the hardware level~\cite{tpmfail}. 
%

\paragraph{Microarchitectural Threats}
Arguably, one of the greatest security threats comes from attacks that target the implementation through side-channels or from hardware vulnerabilities. Such attacks started as a niche exploiting leakages through execution timing, power, and electromagnetic emanations but later evolved to exploit microarchitectural (\Mi ) leakages, e.g. through shared cache and memory subsystems, speculative execution, shared peripherals, etc. \Mi\ threats represent one of the most significant types of vulnerabilities since they can be carried out remotely with software access only. Prime examples of these threats are the early execution timing~\cite{kocher1996timing} and cache attacks~\cite{yarom2014flush+, liu2015last,liu2015last}, and later Meltdown, Spectre~\cite{kocher2019spectre}, and MDS attacks~\cite{moghimi2020medusa,canella2019fallout,vanbulck2020lvi}  which allow an unprivileged user to access privileged memory space breaking isolation mechanisms such as memory space isolation across processes, cores, browsers tabs and even virtual machines hosted on shared cloud instances. Active attacks, e.g. Rowhammer, have also proven effective in recovering sensitive information~\cite{kwong2020rambleed} and~\cite{mus2023jolt,adiletta2023mayhem}. While numerous practical countermeasures were proposed and implemented, there remains a massive attack surface unexplored. Indeed, 5 years after Meltdown was mitigated (August 2023), a new transient execution vulnerability, Downfall~\cite{moghimi2023downfall}, was discovered that exploits speculative data gathering and allows Meltdown-style data leakage and even injection across threads. 

\paragraph{Lack of Access to Design Internals}
A significant factor contributing to the difficulty of evaluating the security of large-scale computer systems is that design details are rarely disclosed. Given only superficial interface definitions, researchers are forced to reverse engineering and black box analysis. While companies have access to the internals of their system, it is hard to argue that they are aware of their own designs either due to third-party IPs, mobility of engineers, and silos isolating their engineering teams from each other. IPs are orphaned with little superficial information surviving after only a few years of breaking institutional memory. These factors combined pose a great danger for \Mi\ security. 

The primary goal of the proposed work is to answer the following question: \textit{Can we use AI to automatically find brand-new vulnerabilities?}
In practical terms, can we build an AI agent that can discover the next Meltdown or Spectre vulnerabilities? Currently, there are intense efforts in the cybersecurity research community to deploy AI tools to scan Open Source Software (OSS) for known vulnerabilities, e.g. for detection in \Mi\ we have~\cite{doychev2015cacheaudit,jan2018microwalk,jan2022microwalk,cauligi2020constant,wang2019kleespectre,guarnieri2020spectector} and for patching~\cite{gupta2017deepfix,yasunaga2021break,tarlow2020learning,pearce2023examining,wu2023effective,garg2023rapgen} and~\cite{tol2023zeroleak}.

We take on a more challenging problem and investigate how we can build an AI Agent that constantly searches the target platform for brand new \Mi\ vulnerabilities. In a way, such an ability would bring true scalability and a tipping point since, if granted, we could surpass human abilities by creating as many AI Agents as we want by just throwing more cycles at the problem. In the hands of software/hardware vendors, such a tool would allow us to address vulnerabilities early on before the software advances deeper in the deployment pipeline. What is missing is the know-how to put such a system together i.e. a tool that can constantly analyze a hardware/software stack under popular configurations, identify and report found vulnerabilities, articulating cause and effect and severity of the vulnerability. In this work, we take inspiration from cybersecurity researchers on how they came up with new vulnerabilities:

\paragraph{Randomization} There is a healthy dose of manual or automated trial and error in discovering new vulnerabilities. In \Mi\, security fuzzing has become an indispensable tool to test randomized attack vectors and thereby identify or generate improved versions of vulnerabilities. For instance, Oleksenko et al.\cite{oleksenko2019specfuzz} developed SpecFuzz to test for speculative execution vulnerabilities. The tool combines dynamic simulation with conventional fuzzing for the identification of potential Spectre vulnerabilities. Another example is Transyther~\cite{moghimi2020medusa} , a mutational fuzzing tool that generates Meltdown variants and tests them to discover leaks. Transyther found a previously unknown transient execution attack through the word combining buffer in Intel CPUs~\cite{moghimi2020medusa}. In~\cite{jattke2022blacksmith}, Jattke et al. use fuzzing to discover non-uniform hammering patterns to make Rowhammer fault injection viable in a large class of DRAM devices. While effective, fuzzing, as currently practiced in \Mi\ security, only works in small domains and fails to scale to cover larger domains to discover new vulnerabilities. Indeed, SpecFuzz for Spectre v1 is only able to Spectre gadgets, and Transyther discovered the Medusa vulnerability since it is reachable with mild randomization from Meltdown variants.

The discovery of the timing channel by Kocher~\cite{kocher1996timing} led to the discovery of cache-timing attacks~\cite{osvik2006cache}. Similarly, sharing in Branch Prediction Units (BPUs) led to the exploitation of secret dependent branching behavior to recover leakages~\cite{aciiccmez2006predicting}. These attacks led to \Mi\ Covert Channels that may be used intentionally to exfiltrate data, e.g. by signaling via cache access patterns and break isolation mechanisms. Covert-channels were first used by many researchers as an initial demonstration of the existence of a side-channel, with the channel rate providing a measure for the level of the leakage. Covert channels and manipulations in BPUs, in turn, became enablers for Transient Execution Attacks such as Meltdown, Spectre, and later MDS attacks. Further, the recent work~\cite{moghimi2023downfall} uses the Meltdown style data leakage and the LVI style~\cite{vanbulck2020lvi}  data injection mechanisms in the context of SIMD instructions to discover new vulnerabilities.

The x86 instruction set is a complex architecture that supports thousands of instructions, registers, and addressing modes, with each microarchitecture adding layers of optimizations for performance and efficiency. These optimizations, while beneficial, introduce complexities that can hide vulnerabilities, as seen with exploits like Meltdown and Spectre, which exploit unexpected microarchitectural behavior to expose sensitive data.
Traditional testing methods like random fuzzing are inadequate due to the vast number of instruction combinations and the specific, rare conditions that often trigger vulnerabilities. Complex features like out-of-order and speculative execution increase both performance and the difficulty of detecting flaws, making the discovery of microarchitectural vulnerabilities challenging.

An effective approach involves intelligent, feedback-based testing, where processor behavior under different conditions guides the search for vulnerabilities. This approach allows testing to focus on high-priority areas, improving efficiency and effectiveness. Feedback mechanisms can also adapt to new microarchitectures, adjusting their methods for each processor generation, an essential feature given the rapid evolution of hardware designs.
Machine Learning (ML) enhances this feedback-driven approach by identifying patterns in cache or power usage that indicate potential vulnerabilities. Over time, ML models improve, enabling more systematic and scalable vulnerability discovery across diverse processor designs.
RL further advances this approach, using a reward-based system to optimize instruction space exploration. RL agents prioritize instruction sequences that reveal anomalies, efficiently balancing exploration with exploiting known vulnerabilities, making them suitable for evolving architectures.

In summary, random fuzzing alone is insufficient for discovering vulnerabilities in modern x86 microarchitectures. Integrating feedback mechanisms with RL allows a more targeted, adaptable, and effective approach, essential for uncovering hidden vulnerabilities and maintaining security in rapidly advancing processor designs.

In this work, we make the following contributions:
\begin{enumerate}[nosep]
    \item We propose a novel approach to discovering microarchitectural vulnerabilities using RL.
    
    \item We develop a custom RL environment that simulates the execution of x86 instructions on a microarchitecture, allowing the agent to explore the instruction space.
    
    \item We find new transient execution based leakage mechanisms on Intel Skylake-X and Raptor Lake microarchitectures based on masked FP exceptions and MME/x87 transitions demonstrating the effectiveness of the RL agent in discovering vulnerabilities.

\end{enumerate}

\section{Related Works}

\Mi vulnerability discovery has attracted significant attention, leading to the development of several tools and methodologies aimed at exposing speculative execution and side-channel vulnerabilities. 
Osiris~\cite{weber2021osiris} introduces a fuzzing-based framework that automates the discovery of timing-based \Mi side channels by using an instruction-sequence triple notation: reset instruction (setting the \Mi component to a known state), a trigger instruction (modifying the state based on secret-dependent operations), and a measurement instruction (extracting the secret by timing differences).
Transynther~\cite{moghimi2020medusa, moghimi2023downfall} automates exploring Meltdown-type attacks by synthesizing binarizes based on the known attack patterns. For the classification and root cause analysis of the generated attacks, Transynther uses performance counters and \Mi ``buffer grooming'' technique.
AutoCAT~\cite{luo2022autocat} automates the discovery of cache-based side-channel attacks on unknown cache structures using RL.
Several studies also focus on using hardware performance counters to detect speculative execution issues. For example,~\cite{ragab2021rage, oleksenko2023hide} use performance counters to monitor mis-speculation behavior. 
More recently, \cite{chakraborty2024shesha} proposed a particle swarm optimization based algorithm to discover unknown transient paths. Their main assumption is different instruction sets do not interfere with each other do not share the same resources, therefore, they can be analysed independently. In this dissertation, we show that this assumption limits the exploration of the instruction space and combining different instruction sets can lead to new mechanisms of transient execution.

Although, these tools have shown promise in detecting \Mi vulnerabities, they are limited in their ability to efficiently explore the large instruction space and the complex interactions between different instructions.


\section{Background}
\subsection{Microarchitectural Attacks}
\subsubsection{Cache Timing Side Channel Attacks}

The state of the shared cache can be observed to detect the memory access patterns. Over the past years, different techniques have been developed to extract sensible data by using cache timing as a side-channel attack.

Flush+Reload~\cite{yarom2014flush+} leverages the Last-Level Cache (L3 cache) to monitor memory access patterns in shared pages. While it does not require the attacker and victim to share the same execution core, it flushes a potential victim address from the cache, and then measures the reload time if the target address is accessed. EVICT+RELOAD~\cite{gruss2015cache} is another work where an eviction technique is used when cache flushing is not available. Prime+Probe~\cite{liu2015last} exploits the eviction sets to detect access patterns and it does not require shared memory between attacker and victim.
The Flush+Flush attack~\cite{gruss2016flush+} exploits the timing variations of the \texttt{clflush}. The Evict+Time attack~\cite{osvik2006cache} uses timing differences between cache hits and misses to infer cache state, allowing an attacker to detect cache misses and deduce access patterns. 

Although more advanced cache side-channel attacks~\cite{irazoqui2015s, disselkoen2017prime+, purnal2021prime+}, in this work, we use Flush+Reload for its simplicity and effectiveness.


\subsubsection{Transient Execution Attacks}

Transient execution attacks exploit speculative and out-of-order execution in CPUs to access restricted data temporarily, leaving traces in the cache that attackers can analyze. 

Spectre attacks~\cite{kocher2019spectre} exploit speculative execution and branch prediction in modern processors to leak confidential information across security boundaries. By inducing speculative operations that bypass normal execution flow, attackers can access sensitive memory and registers, creating side-channel leaks. NetSpectre~\cite{schwarz2019netspectre} is the first remote variant of the Spectre attack, extending its reach beyond local code execution. NetSpectre marks a significant shift from local to remote attacks, making Spectre a threat even to systems where no attacker-controlled code is executed, including cloud environments.

SgxPectre attack~\cite{chen2019sgxpectre} exploits CPU vulnerabilities to compromise the confidentiality and integrity of SGX enclaves. By manipulating branch prediction from outside the enclave, attackers can temporarily alter the enclave's control flow, producing cache-state changes that reveal sensitive information within the enclave.

Meltdown~\cite{Lipp2018meltdown} bypasses memory isolation by exploiting out-of-order execution in modern processors to access protected kernel memory. This enables attackers to read memory from other processes or virtual machines without permission, posing a severe risk to millions of users. Foreshadow~\cite{vanbulck2018foreshadow} is a microarchitectural attack exploiting speculative execution flaws in Intel processors to breach SGX security. Without needing kernel access or assumptions about enclave code, Foreshadow leaks enclave secrets from the CPU cache.

Rogue In-flight Data Load (RIDL)~\cite{van2019ridl} is a speculative execution attack that leaks data across address spaces and privilege boundaries. RIDL retrieves in-flight data directly from CPU components without relying on cache or translation structures, making it uniquely invasive and effective. Fallout~\cite{canella2019fallout} reveals that Meltdown-like attacks remain feasible on newer CPUs that are supposedly immune to Meltdown due to hardware fixes. By examining the behavior of the store buffer, they uncover vulnerabilities.

ZombieLoad~\cite{Schwarz2019ZombieLoad} is a Meltdown-type attack that exploits a vulnerability in the processor’s fill-buffer logic to leak data across logical CPU cores, even on CPUs with hardware mitigations against Meltdown and MDS. ZombieLoad uses faulting load instructions to transiently access unauthorized data in the fill buffer. Load Value Injection (LVI)~\cite{van2020lvi} is a technique that generalizes injection-based attacks to the memory hierarchy by injecting attacker-controlled values into a victim’s transient execution. Downfall~\cite{moghimi2023downfall} is a new class of transient execution attacks that exploit the gather instruction on x86 CPUs to leak sensitive data across security boundaries, including user-kernel, process, and virtual machine isolation, as well as trusted execution environments.

\subsection{Reinforcement Learning}

In RL, the objective is for an agent to learn a policy $\pi_\theta(a|s)$, parameterized by $\theta$, which maximizes the expected cumulative reward through its chosen actions in an environment. The policy gradient method~\cite{sutton1999policy} computes the gradient of the expected reward with respect to the policy parameters, allowing the agent to directly update the policy by following the gradient. Formally, the objective function $J(\theta)$ is defined as:
%
\[
J(\theta) = \mathbb{E}_{\pi_\theta} \left[ \sum_{t=0}^{T} r_t \right],
\]
%
where $r_t$ is the reward at time step $t$, and the expectation is over the trajectories induced by the policy $\pi_\theta$. The policy is updated by adjusting $\theta$ in the direction of the gradient $\nabla_\theta J(\theta)$ using gradient ascent. One of the major challenges with vanilla policy gradient methods is the high variance of the gradient estimates, which can lead to unstable learning. Additionally, large updates to the policy parameters $\theta$ can cause dramatic changes to the policy, potentially leading to performance collapse. 
Trust Region Policy Optimization (TRPO)~\cite{schulman2015trpo} was proposed to address this issue by enforcing a constraint on the size of policy updates using a trust region.
TRPO introduces the following constrained optimization problem:
%
{\small
\[
 \max_\theta \mathbb{E}_{\pi_\theta} \left[ \frac{\pi_\theta(a|s)}{\pi_{\theta_\text{old}}(a|s)} \hat{A}(s,a) \right] 
 \quad \mbox{subject to} \quad \mathbb{E}_{s} \left[ D_{\text{KL}} \left( \pi_{\theta_\text{old}} \| \pi_\theta \right) \right] \leq \delta
\]
}
%
where $D_{\text{KL}}$ is the Kullback-Leibler (KL) divergence, $\hat{A}(s,a)$ is the advantage estimate, and $\delta$ is a small positive value controlling the step size. However, TRPO is computationally expensive due to the need for second-order optimization to enforce the KL-divergence constraint.

Proximal Policy Optimization (PPO)~\cite{schulman2017ppo} simplifies TRPO by replacing the hard constraint on policy updates with a penalty or by using a clipped objective function. The key idea behind PPO is to ensure that policy updates are ``proximal'' to the current policy, preventing drastic updates that could lead to instability. 

In this work, we use PPO with \textit{clipped objective}. In this approach, PPO clips the probability ratio $\frac{\pi_\theta(a|s)}{\pi_{\theta_\text{old}}(a|s)}$ to lie within a small interval around 1, preventing large updates. The clipped objective is defined as:
{\small
\[
L^{\text{CLIP}}(\theta) = \mathbb{E} \bigg[ \min \big( r(\theta) \hat{A}(s,a), \text{clip}(r(\theta), 1 - \epsilon, 1 + \epsilon) \hat{A}(s,a) \big) \bigg],
\]
}
%
where $r(\theta) = \frac{\pi_\theta(a|s)}{\pi_{\theta_\text{old}}(a|s)}$ is the probability ratio, and $\epsilon$ is a small hyperparameter that limits how far the policy is allowed to change. By clipping the probability ratio, PPO discourages overly large updates while still allowing for sufficient exploration of the policy space.


\section{Threat Model and Scope}
Our threat model considers scenarios where attacker and victim processes are co-located in shared hardware environments, which expose vulnerabilities to \Mi attacks.
Co-location can manifest in several forms, including but not limited to threads on the same process, processes on the same host and virtual machines on a shared server. These attacks exploit shared \Mi resources to infer sensitive data from victim processes, bypassing traditional memory isolation mechanisms.

We assume the CPU microcode is up-to-date with the latest mitigations, and the software has no bugs and SMT is enabled.
We assume no access to the confidential design details of the processor, limiting our analysis to black-box testing. This restriction reflects the real-world scenario where attackers must rely on external observations and performance counters to reverse-engineer the processor's internal behavior.

Although we have not seen example of such an exploit in real-life yet, if unmitigated, these attacks can lead to significant data breaches, including the extraction of cryptographic keys and other sensitive information.
In this work, we focus on discovering \Mi vulnerabilities using reinforcement learning and we focus on the following questions: 
\medskip

\noindent
\textbf{Q1.} How can we design an RL framework that efficiently explores the \Mi space? \\

\noindent \textbf{Q2.} Can RL discover unknown \Mi vulnerabilities? \\

\noindent \textbf{Q3.} What are the challenges and limitations of using RL for \Mi vulnerability discovery?

\section{Our RL Framework}

\begin{figure*}
    \centering
    \includegraphics[width=0.8\textwidth]{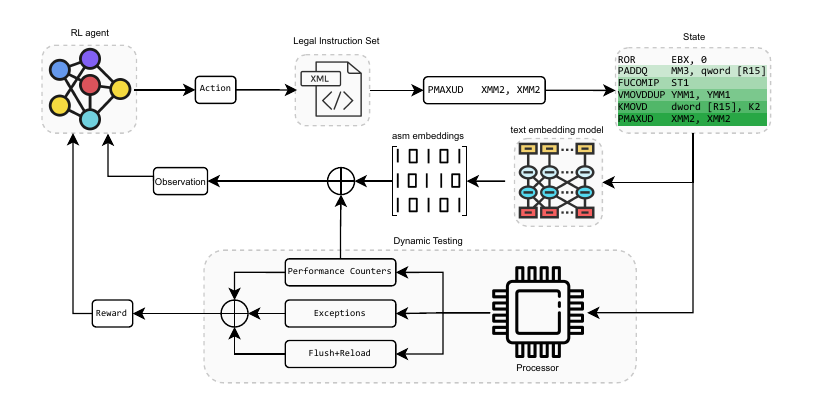}
    \caption{\label{fig:rl_framework}Overview of the RL framework for \Mi vulnerability analysis; $\bigoplus$ before the \textit{Observation} denotes concatenation. $\bigoplus$ before \textit{Reward} represents Equation~\ref{eq:reward}.}
\end{figure*}

In this section, we introduce our RL framework designed for \Mi vulnerability analysis.
Automated analysis of \Mi vulnerabilities poses the following challanges some of which were also identified in earlier works~\cite{weber2021osiris,moghimi2020medusa,chakraborty2024shesha}:

\begin{itemize}[nosep, leftmargin=*]
\item \textbf{C1.} Modern processor designs are complex and their instructions sets are large. Exhaustive search in the instruction space is infeasible. 
\item \textbf{C2.} Mapping an instruction sequence to a certain \Mi vulnerability is non-trivial and requires expert knowledge. 
\item \textbf{C3.} The environment is high-dimensional and non-linear and the system state is only partially observable. 
\end{itemize}


Earlier works attempted to solve \textbf{C1} by either limiting the type of instructions~\cite{chakraborty2024shesha,oleksenko2023hide} or limiting the length of the instruction sequence~\cite{weber2021osiris}. In this work, we propose a novel approach to address this challenge by leveraging RL to guide the search for \Mi vulnerabilities. Our framework is designed to efficiently explore the instruction space, learn the optimal policy for selecting instruction sequences, repoduce known vulnerability and, if exists, discover unknown vulnerabilities. The framework is illustrated in Figure~\ref{fig:rl_framework} and consists of the following components:

\subsection{Environment} We build a custom environment based on the underlying CPU model. The environment represents a black-box model of the CPU microarchitecture, where the agent can only interact with the CPU through the instruction sequences. It takes the instruction sequences generated by the RL agent, executes them on the CPU, and returns an observation and a reward.
It also updates the state after every action taken. 


At the start of each episode, the environment initializes by resetting the instruction state and clearing the performance counter readings. A reset function is triggered at the beginning of each new episode to ensure the agent starts with a fresh state.
All sequences, performance metrics, and detected byte leakages are logged for post-training analysis. The logged data aids in identifying patterns or characteristics in sequences that lead to vulnerabilities and provides insights into the agent's decision-making process.

\subsection{RL Agent} The RL agent is a multi-layer perceptron (MLP) that generates actions based on observations given by the environment. In this case, the agent's goal is to select an instruction that will be appended to the instruction sequence. The agent is trained using the PPO algorithm. The goal of the agent is to maximize the reward signal by selecting the best sequence of actions and eventually trigger \Mi vulnerabilities.

\subsection{Action Space} 
We define an action as the selection of an assembly instruction from the instruction set. To map the discrete actions to actual assembly instructions, we use~\cite{abel19a}.
The action space is constrained to instructions that are supported by the CPU under test and documented by the vendor. This constraint helps the agent focus on relevant instructions that exist in the real-world programs. Since some of the instruction extensions has large number of instructions and operand variety, (e.g. AVX-512), we construct the action space hierarchically. For example, we first select the instruction set (e.g. AVX-512), then the instruction (e.g. \texttt{VMOVDDUP}), and finally the operands (e.g. \texttt{XMM0}, \texttt{XMM1}). This hierarchical structure helps prevent larger instructions sets dominating the smaller ones since the agent will initially randomly select insturction during the exploration phase. To handle the difference in the number of instructions in each set, we use map different actions to the same insturction or operands using them modulo operation.
For instance, if the maximum number of instructions in a set is 10 but the model selected the $12^{th}$ instruction, we map it to the $12 \bmod 10 = 2^{nd}$ instruction in the selected set.


\subsection{State} Eventhough, there are more variables that affect the CPU state other than just the input instruction sequence, such as, cache content, internal buffers, registers, etc., we simplify the state representation to only the instruction sequence. The impact of other factors that affects the CPU state can be minimize by running the same instruction sequence multiple times until the real state becomes stable, which is a common practice in \Mi attacks~\cite{yarom2014flush+, liu2015last}. After each action, the generated assembly instruction is added to the current state.

\subsection{Observation} 
Since we do not have access to hardware debug interface, we cannot directly observe the entire state of the CPU. Therefore, it is a \textit{partially observable} environment and the observation can only capture a subset of the environment state as it is mentioned in~\textbf{C3}. We tackle this challange by designing an observation space that consists of a static and a dynamic part.

The static part of the observation is the generated instruction sequence. Similar to the earlier works~\cite{tol2021fastspec,mankowitz2023alphadev}, we use embeddings to convert the instruction sequence into high-dimensional fixed-size vectors using a pre-trained LLM. Embeddings capture the patterns in the assembly code so that the agent understand the structural and functional dependencies between instructions. Before inclusion in the observation space, embeddings undergo normalization to ensure consistency in data scales. 

The dynamic part of the observation is the hardware performance counters. Vendors give access to low-level monitoring of the CPU events such that developers can identify bottlenecks in their applicaitons and optimize the performance. In this work, we use the performance counters to partially capture the CPU state. For  measurement, we embed instruction sequences in a template assembly file, ensuring valid memory addresses in \texttt{R15} register to prevent segmentation faults. General-purpose registers are preserved on the stack to avoid unintended corruption. Each sequence is executed multiple times to minimize noise.

\begin{table}[h!]
    \centering
    \footnotesize
    \begin{tabular}{lll}
    \toprule
    \textbf{HW Performance Event Name} \\
    \midrule
    UOPS\_ISSUED.ANY  \\
    UOPS\_RETIRED.RETIRE\_SLOTS \\
    INT\_MISC.RECOVERY\_CYCLES\_ANY  \\
    MACHINE\_CLEARS.COUNT \\
    MACHINE\_CLEARS.SMC \\
    MACHINE\_CLEARS.MEMORY\_ORDERING \\
    FP\_ASSIST.ANY$^a$ \\
    OTHER\_ASSISTS.ANY$^b$ \\
    CPU\_CLK\_UNHALTED.ONE\_THREAD\_ACTIVE \\
    CPU\_CLK\_UNHALTED.THREAD \\
    HLE\_RETIRED.ABORTED\_UNFRIENDLY$^c$ \\
    HW\_INTERRUPTS.RECEIVED$^d$ \\
    \bottomrule
    \end{tabular}
    \caption{List of used CPU performance events available in Skylake-X. In Raptor Lake, corresponding events: $a$-- ASSISTS.FP, $b$--ASSISTS.ANY, $c$--N/A, $d$--N/A.}\label{tab:cpu_events}
\end{table}

We use the performance counters listed in Table~\ref{tab:cpu_events}. The explanation of these counters are given in the Appendix~\ref{sec:hpc_descriptions}. 
These counters are selected based on their relevance to speculative execution vulnerabilities shown by previous research~\cite{ragab2021rage,oleksenko2023hide,chakraborty2024shesha} as well as Intel's performance monitoring tools~\cite{intel_perfmon}.

\subsection{Reward Function}\label{sec:reward}

The reward function is often seen as the most critical component of the RL frameworks since it steers the agent behavior. We address the challenge \textbf{C2} by carefully designing the reward function.

The instruction sequences selected by the agent are executed on the CPU, and the CPU's behavior is monitored using hardware performance counters. The counters provide feedback on the speculative execution and microarchitectural effects of the instructions.

The reward function evaluates the performance counter data collected during instruction execution. It assigns rewards based on the presence of speculative execution anomalies, deviations from expected behavior, or other indicators of potential vulnerabilities. 
    
%
\begin{equation}\label{eq:reward}
    \begin{aligned}
        \text{Reward} &= \frac{\textit{bad speculation} + \textit{observed byte leakage} }{\textit{instruction count}}
    \end{aligned}
\end{equation}
Equation~\ref{eq:reward} shows a simplified version of the reward function we use for training the RL agent. The reward is calculated by dividing the sum of bad speculation and observed byte leakage by the number of instructions in the sequence. 
The final reward value is capped at 100 if $\textit{observed byte leakage} = 0$ and 500 if $\textit{observed byte leakage}>0$ to prevent excessive rewards, promoting stable training. The numbers were decided based on the empirical results.

\paragraph{Testing for Bad Speculation}
According to Intel's documentation~\cite{intel_vtune_2023}, ``\textit{bad speculation}'' typically results from branch mispredictions, machine clears and, in rare cases, self-modifying code. 
It occurs when a processor fills the instruction pipeline with incorrect operations due to mispredictions. This process leads to wasted cycles, as speculative micro-operations (uops) are discarded if predictions are incorrect, forcing the processor to recover and restart. Although bad speculation is primarily a concern for performance, it also has important security implications.
Microarchitectural attacks exploit transient states created by bad speculation. During speculation, the CPU may access sensitive data or load it into the cache, even though the operations will eventually be discarded. These transient states, particularly in cache memory, create opportunities for attackers to infer sensitive data—such as encryption keys—by analyzing cache behaviors and measuring access times. 

Intel's formula for quantitatively measurement of \textit{bad speculation} for a CPU thread is
\begin{multline}
    \text{Bad\ Speculation} = \text{UOPS\_ISSUED.ANY} \\
    - \text{UOPS\_RETIRED.RETIRE\_SLOTS} \\
    + \left( 4 \times \text{INT\_MISC.RECOVERY\_CYCLES} \right)
\end{multline}
which also what we use in our reward function. 

If there is an exception detected during the performance counter tests, we terminate the episode, set the reward to -10 and reset the state. We select this number arbitrarily to differentiate between instruction sequences with no bad speculation vs instruction sequences that do not execute at all. Negative reward discourages the agent from generating exceptions. Note that, handling the exceptions is also possible, but it complicates the reward calculation. Therefore, we leave it for future work.

\paragraph{Testing for Observable Byte Leakage}

\begin{figure}[h]
    \vspace{-4mm}
        \centering
        \includegraphics[width=\columnwidth]{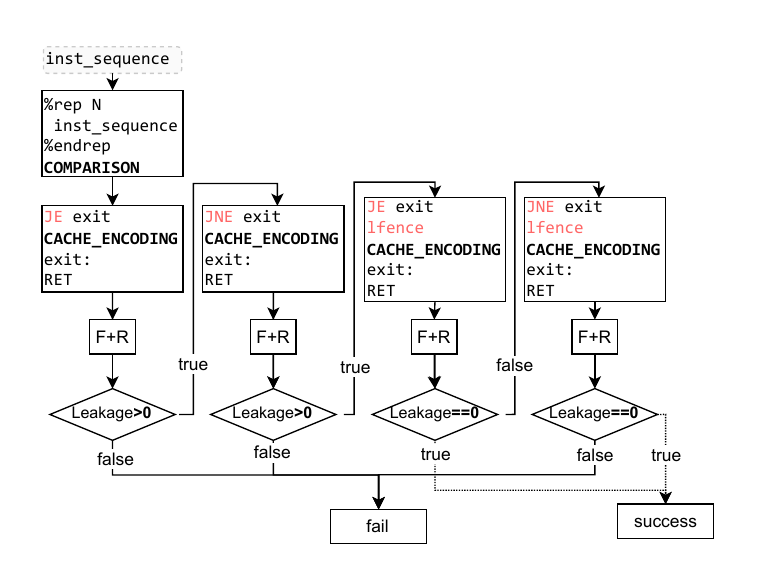}
        \caption{Test flow for detecting observable byte leakage.}\label{fig:leakage_test}
    \end{figure}
    
If the performance counter tests executes successfully, we check if the generated instruction sequence results in observable byte leakage due to speculative execution.
Our testing flow for detecting observable byte leakage is shown in Figure~\ref{fig:leakage_test}.

We, first, place the instruction sequence in a template assembly file and run it \texttt{N} times using \texttt{rep} directive. 
Similar to performance counter tests, we use predefined addresses for memory operands and preserve the contents of the general purpose registers in the stack.
Then, we execute a comparison operation based on the instructions types and registers used in the generated sequence. If there are multiple types of registers used in the sequence, we select a different comparison instruction specific to that register type. We repeat the test flow for each register type used in the sequence. This way we avoid false negatives due to the register type mismatch.
After the comparison, we execute a conditional branch instruction (jump if equal--\texttt{JE}) which is followed by cache accesses to an array that encode a predefined sequence of bytes to the cache state. We then measure the access time to the array using Flush+Reload~\cite{yarom2014flush+} to decode the bytes and check if it how much it matches with the encoded bytes.
If there is any match, we repeat the same test, this time with the opposite branch condition (jump if not equal--\texttt{JNE}). If there is a match in this case, we consider it as an observable byte leakage. 

Note that, most of the generated seqences fail either in the first or second step of the leakage test. For the remaining sequences that passed the first two tests, we run the same two test after inserting \texttt{lfence} before the branch instruction. If the leakage disappears after adding \texttt{lfence}, we consider it as a successful sequence that causes observable byte leakage through bad speculation. Note that, unlike Spectre-BHT~\cite{kocher2019spectre}, we do not train the branch predictor in the test flow so the root cause of the bad speculation would not be the branch mispredictions unless the generated sequence has the branch predictor training itself using branch instructions.

We repeat the test flow for each register type used in the sequence. This way we avoid false negatives due to the register type mismatch.
The number of successfully decoded bytes are fed into the reward function as the observable byte leakage. Since the byte leakage is a more direct signal of the vulnerability, we assign a higher weight to it in the reward function.
If an exception is detected at this stage, the environment resets to a safe state, logs the exception. Only the byte leakage part of the reward is set as zero, yet the bad speculation part is calculated as usual.

\section{Experiments}

\begin{figure}
    \centering
    \includegraphics[width=\linewidth]{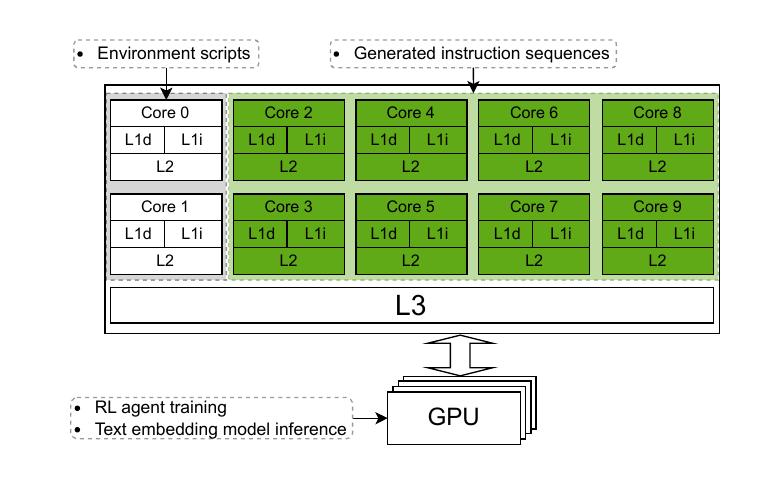}
    \caption{Experiment Setup on Skylake-X. Each physical core shown in green was allocated to a single instruction sequence at a time.}\label{fig:experiment_setup}
\end{figure}

\paragraph{Experiment Setup}
We run the experiments on two systems with different \Mi. The first system has an Intel Core i9--7900X CPU with a Skylake-X \Mi. The OS running on the system is Ubuntu 22.04.5 LTS with the Linux v6.5.0--44-generic. We use glib v2.72.4, nasm v2.15.05, and gcc v11.4.0 for compiling and testing the generated assembly files; PyTorch v2.2.1, Stable Baselines3 v2.2.1 and Gymnasium v0.29.1 for custom RL environment and training the RL agent. 
The RL agent training and the local inference for the text embedding model are done on the GPU clusters with an NVIDIA TITAN Xp, GeForce GTX TITAN X, and two GeForce GTX 1080Ti. The overview of the experiment setup used in the first system is illustrated in Figure~\ref{fig:experiment_setup}.

The second system has an Intel i9-14900K CPU with a Raptor Lake \Mi, Ubuntu 24.04.1 LTS with Linux v6.8.0-51-generic, glib v2.80.0, nasm v2.16.01, gcc v13.3.0. Pytorch v2.4.1+cu121, Stable Baselines3 v2.3.2, and Gymnasium v0.29.1 are used for the same purposes as the first system. The RL agent training is done on NVIDIA GeForce RTX 4090 GPU.


For remote inference, we use OpenAI's  \texttt{text-embedding-3-small} model through the API access, which allows us to use the parallel core testing since it does not require local GPU memory. Parallelizing the framework across multiple CPU cores enables multiple sequences to be evaluated simultaneously and increases the training speed. Although the last level cache is shared among the cores, each process accesses its own distinct memory region, which does not include any shared libraries or data. Therefore, the cache interference among the processes is minimal.

For local inference, we used the NV-Embed-v2~\cite{lee2024nv,moreira2024nv} embedding model. However, we observed that it does not accelerate the training overall since the GPU memory is not enough to enable parallel core testing. Therefore, the results presented in this paper are based on the text-embedding-3-small model.

In the Skylake-X system, after filtering all illegal instructions from~\cite{abel19a}, we are left with 12598 instructions that belong to 74 sets. The largest set has 2192 instructions, and the maximum number of possible operands per instruction is 7. These numbers determine the size of the action space for the RL agent. 
In the Raptor Lake system, 3996 instructions that belong to 72 sets were left after filtering. The largest set has 468 instructions, and the maximum number of operands per instruction is 7.

For the agent training, we enable all available kernel mitigations against CPU vulnerabilities.
Cumulatively, we collected ~27 days' worth of data from Skylake-X and ~20 days' worth of data from Raptor Lake. In the longest runs, we trained an RL agent for ~4.5 days and ~10.5 days for Skylake-X and Raptor Lake, respectively.
In total, the cost for embedding model inference through API calls was 2.46 USD.

\begin{figure}
    \centering
    \includegraphics[width=\columnwidth]{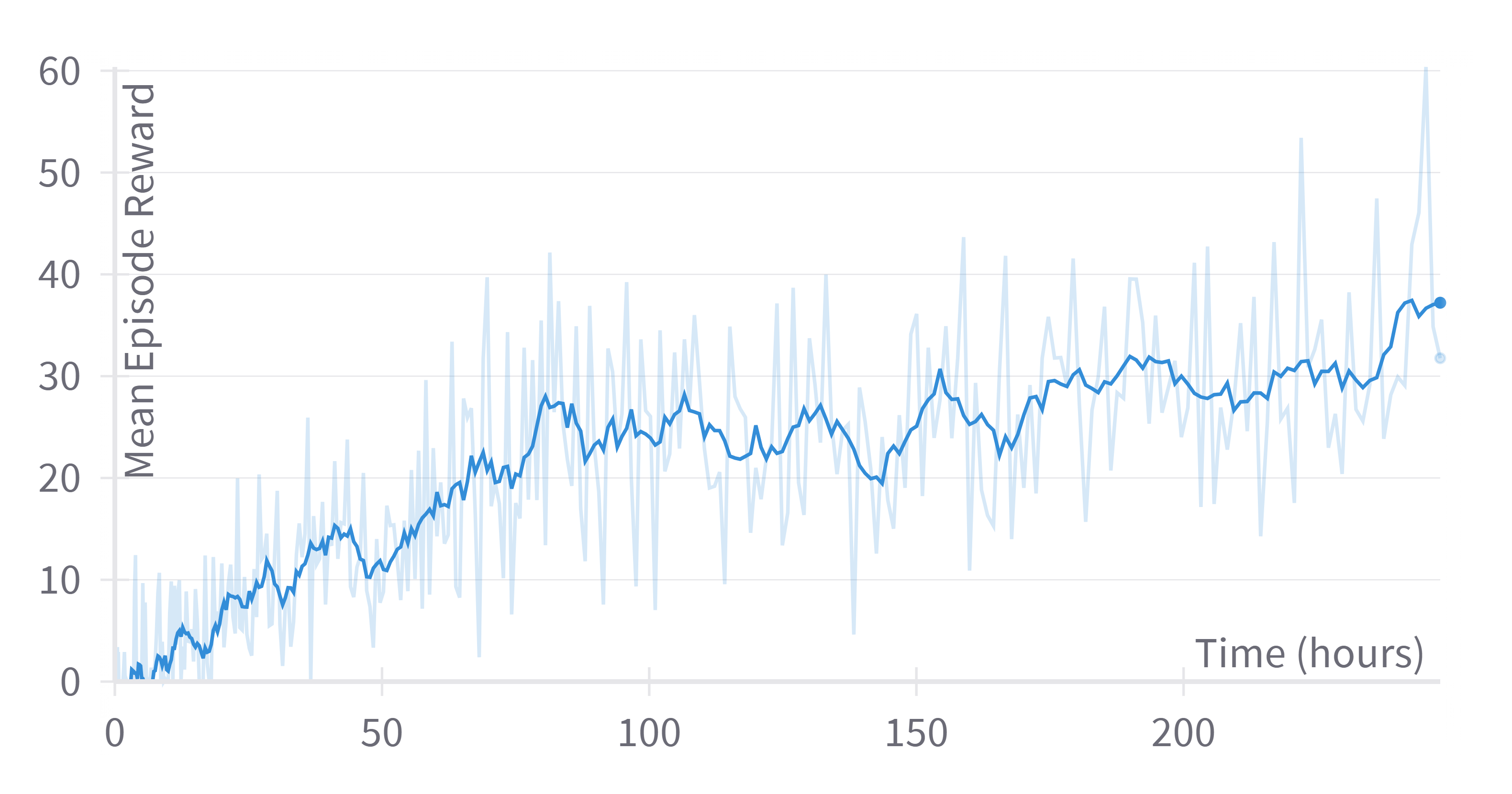}
    \caption{The increase of the average reward per episode during the ~10 days of agent training in Raptor Lake. An episode corresponds to the largest instruction sequence the agent can generate from scratch. The darker line shows the running mean.}
    \label{fig:reward_plot}
\end{figure}

During the RL agent training, we observe that the average reward of each episode increases over time, as shown in Figure~\ref{fig:reward_plot}. The steady increase in reward indicates that the RL agent was able to successfully explore the x86 instruction set while exploiting the knowledge we provide through the reward signal.

\begin{figure}
    \centering
    \includegraphics[width=\columnwidth]{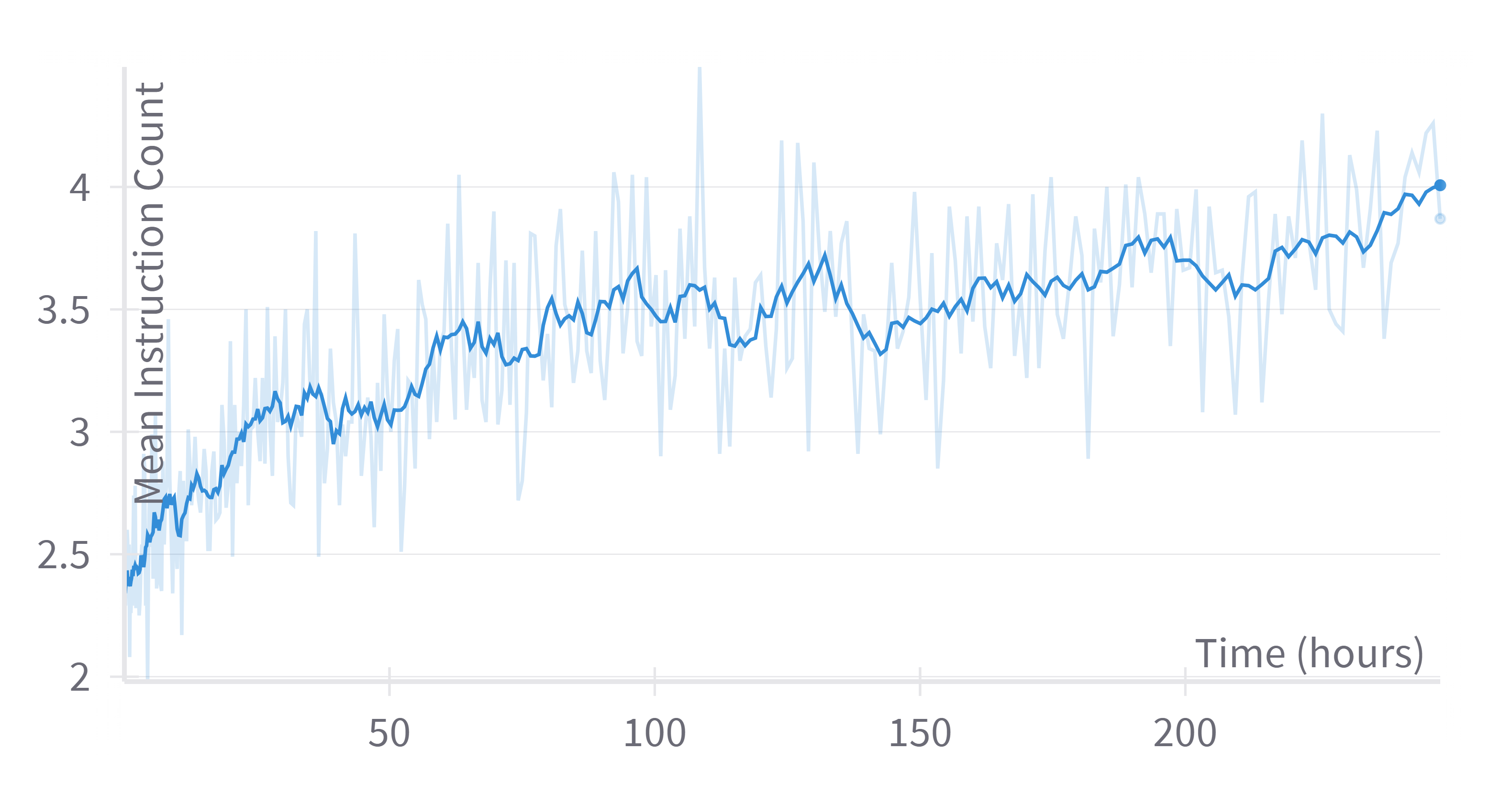}
    \caption{The increase of the average length of generated assembly sequences during the $\sim$10 days of agent training in Raptor Lake. The darker line shows the running mean.}
    \label{fig:episode_length}
\end{figure}

At every episode, the agent starts building an instruction sequence from scratch and keeps adding a new instruction until the instruction limit is reached or the sequence gives an exception with the last added instruction. Over time, we observe that the average length of an instruction sequence during an episode increases during the agent training, as shown in Figure~\ref{fig:episode_length}. Increasing average length over time indicates the agent was able to learn to select combinations of instructions to avoid exceptions.
In our experiments, the maximum sequence length was chosen as 10 assembly instructions, after which the agent starts building a new sequence.

\section{Discovered Transient Execution Vulnerabilities}

\begin{figure}
    \centering
    \includegraphics[width=\linewidth]{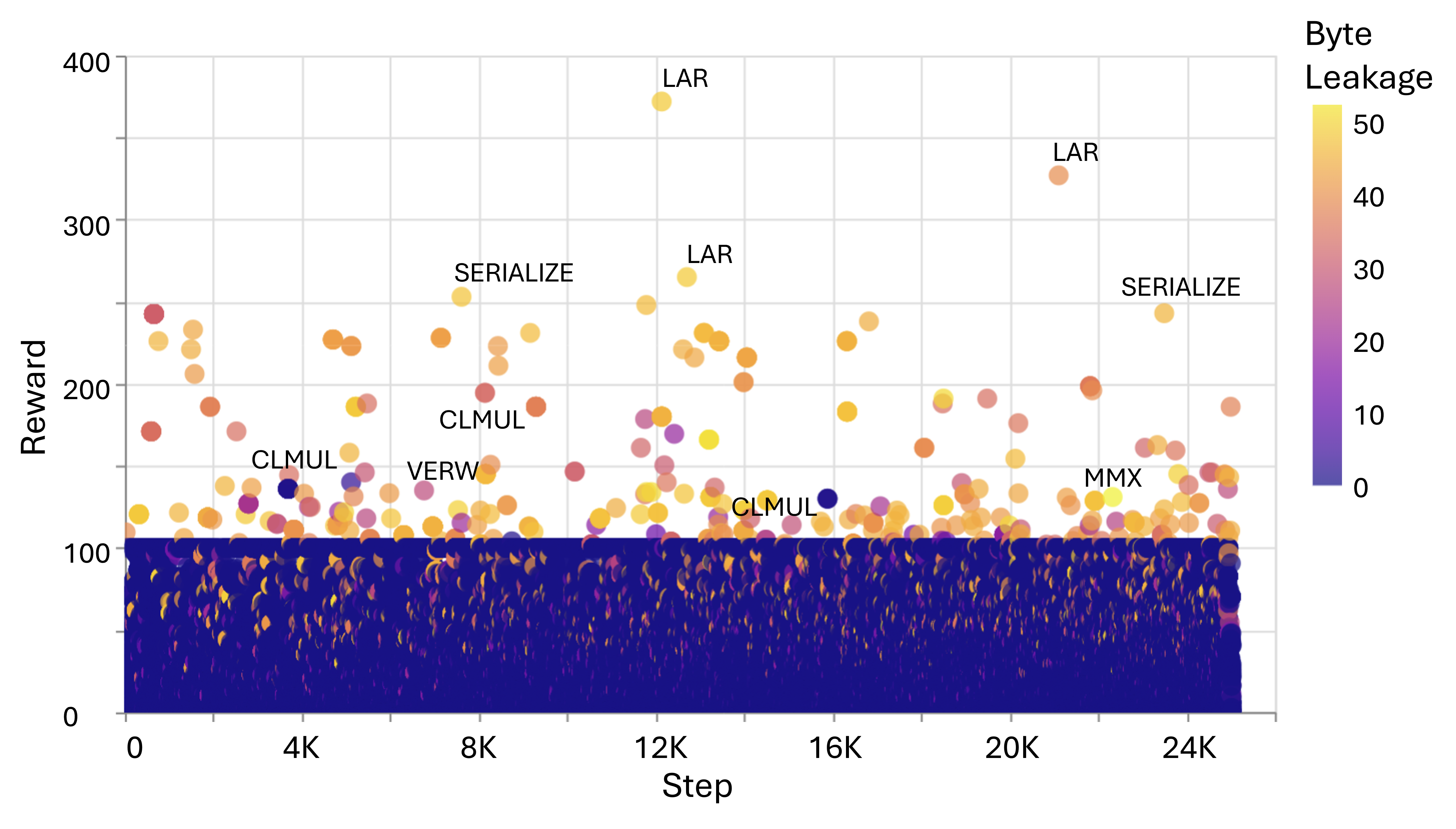}
    \caption{Visualization generated instruction sequences across time, with higher byte leakage marked with lighter colors. Note that, the discovered transient execution mechanisms stand out as it has higher reward compared to the remaining ones.}

    \label{fig:scatter}
\end{figure}

When the agent builds an instruction sequence that results in observable transient execution, it is saved for manual analysis with performance counter and data leakage information. Figure~\ref{fig:scatter} shows the visualization of generated instructions over time. Since the sequences with high reward stands out from the rest of the generated snippets, it eases the manual analysis and verification as well. After analyzing the generated data, we were able to identify \textbf{eight} new classes of transient execution mechanisms that had previously not been documented to the best of our knowledge. Note that, each class has many variations among the generated dataset. Yet here we present, the most simplified versions of the each class.

\paragraph{Masked Exceptions}
The studies in \cite{ragab2021rage,chakraborty2024shesha} demonstrated that FP assists due to denormal numbers cause transient execution of the following instructions.
Our RL agent generated instruction sequences that cause observable byte leakage through transient execution without generating any $\mu$code assists, faults, or interrupts. Listing~\ref{lst:masked_fp} shows an example of such an instruction sequence. 
After careful analysis, we noticed that the sequence indeed causes an FP exception, but the exception is masked by the processor, and the program execution is uninterrupted. 
Previous works reported transient execution with page faults, device-not-available~\cite{Lipp2018meltdown,stecklina2018lazyfp,moghimi2020medusa} which requires exception handling and $\mu$code assists such as FP assists~\cite{ragab2021rage,chakraborty2024shesha} which requires specially crafted inputs.
Transient execution through masked FP exceptions has not been previously reported in the literature, which makes it a new discovery for our RL agent.

We observed this behavior on only Skylake-X.

\begin{figure}[h]
    
\begin{lstlisting}[caption=A simplified instruction sequence that triggers masked FP exception due to repeated x87 instruction in line 4. Following cache encoding instructions (line 8-10) get executed speculatively., label={lst:masked_fp}]
generated_assembly_function:
%rep N
    FLD     qword [x]
%endrep
    FCOMI   st0, st1
    JE      exit:
    ; cache encoding
    MOVZX   rax, byte [%rdi]
    SHL     rax, 10
    MOV     rax, qword [rsi+rax]
exit:
    RET
\end{lstlisting}

\end{figure}

\paragraph{Transitions Between MMX and x87}
FP exceptions, by default, are masked and do not cause a trap, and the program continues execution. However, starting from glibc v2.2, it is possible to unmask them using \texttt{feenableexcept} functions from \texttt{fenv.h} library. This function allows the FP exceptions to cause a trap and the program to be interrupted.

After masked exceptions we run another training session with the same configuration but with the \texttt{feenableexcept} function enabling \texttt{FE\_INVALID}, \texttt{FE\_DIVBYZERO}, \texttt{FE\_OVERFLOW}, \texttt{FE\_UNDERFLOW}, and \texttt{FE\_INEXACT} bits of the \texttt{excepts} argument. 

With this configuration, the RL agent was still able to generate instruction sequences that cause observable byte leakage through transient execution without generating any $\mu$code assists, faults, or interrupts. Listing~\ref{lst:mmx_x87} shows an example of such an instruction sequence.

\begin{figure}
    
\begin{lstlisting}[caption=A simplified version of the RL generated assembly instruction sequence that has MMX (line 3) to x87 (line 4) transition. Following cache encoding instructions (line 9-11) get executed speculatively., label={lst:mmx_x87}]
generated_assembly_function:
%rep N
    PSUBQ MM2, [R15]
    FCOMIP ST4
%endrep
    VCMPPD K3, ZMM1, ZMM4, 2
    JNE exit
    ; cache encoding
    MOVZX   rax, byte [%rdi]
    SHL     rax, 10
    MOV     rax, qword [rsi+rax]
    exit:
    RET
    \end{lstlisting}
\end{figure}

After simplifying the instruction sequence, we observed that the transient execution is caused by an FP exception that is generated by the \texttt{FCOMIP} instruction. However, the MMX instruction before the \texttt{FCOMIP} instruction causes the exception to get lost. We use the \texttt{feenableexcept} function to unmask FP exceptions, yet the exception generated in the processor gets cleared by the \texttt{PSUBQ} instruction. Even though the exception is cleared, the following instructions are executed speculatively, and the transient execution is observed. Note that the comparison instruction \texttt{VCMPPD} does not have any dependency on the previous instructions, yet it is still executed speculatively, and removing the AVX instructions from the sequence does not break the transient execution.

In Intel documentations~\cite{intel-sdm}, it is advised that after the MMX instructions, \texttt{EMMS} instruction should be used to clear the FPU state to prevent ``undefined behavior''. We verified that adding an \texttt{EMMS} instruction after the MMX instruction makes the FP exception cause a trap.

\begin{figure}[htp]
    \centering
    \begin{subfigure}[b]{0.45\textwidth}
        \centering
        \includegraphics[width=\textwidth]{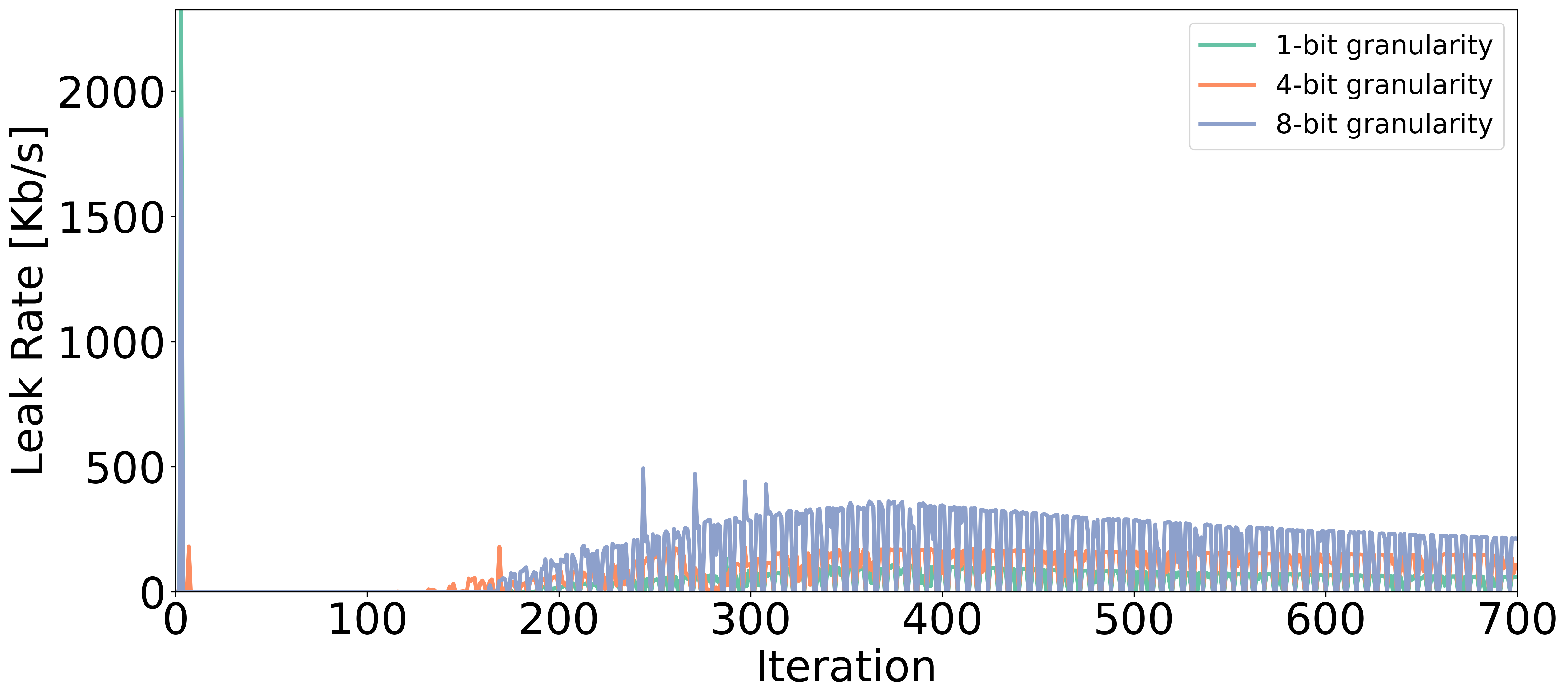}
        \caption{Skylake-X}
        \label{fig:mmx-x87-leak-rate-skylake}
    \end{subfigure}
    \hfill
    \begin{subfigure}[b]{0.45\textwidth}
        \centering
        \includegraphics[width=\textwidth]{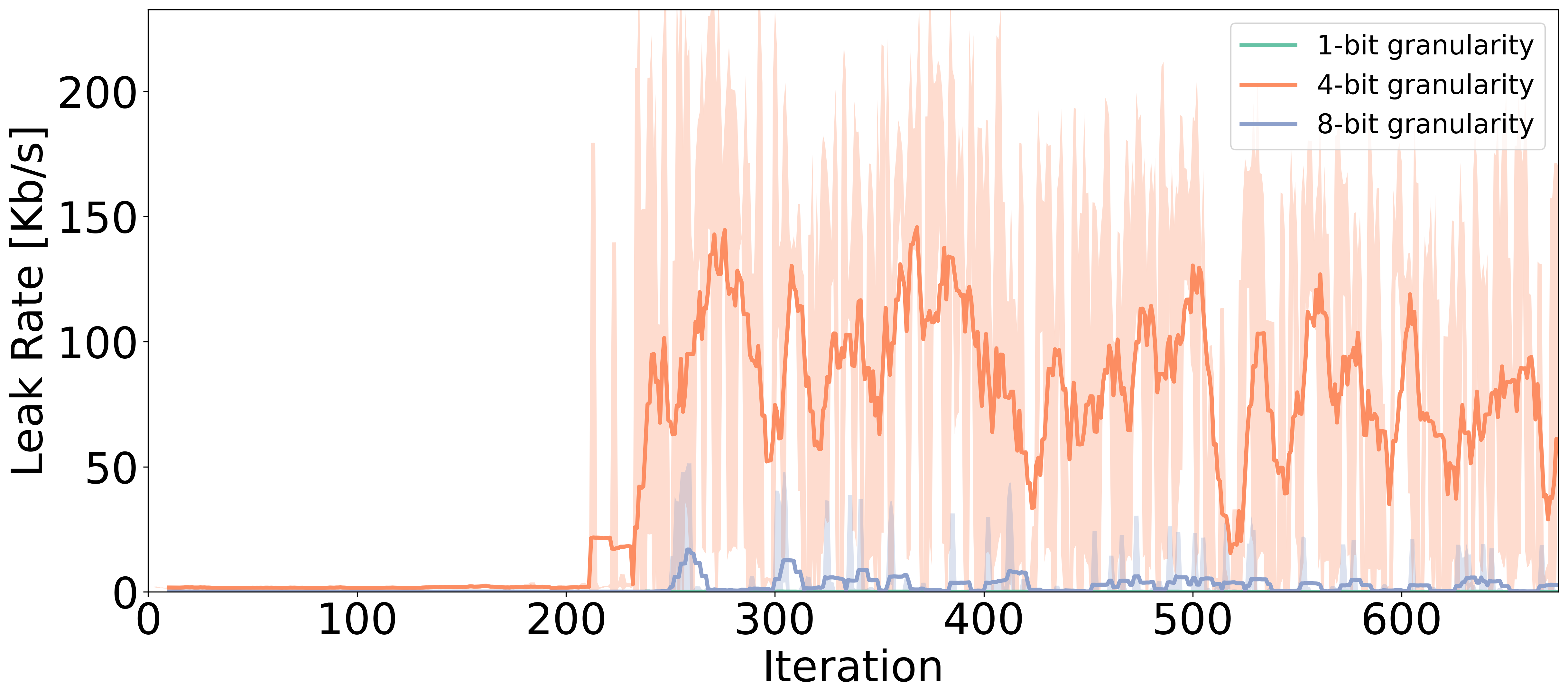}
        \caption{Raptor Lake}
        \label{fig:enter-label}
    \end{subfigure}
    \caption{Leakage rate vs repetitions of the instruction pattern for MMX-x87. Note that in (a) the leakage rates peak when Iteration N=3 for every granularity. In (b), darker lines are running mean and the shades are rolling variance around the mean.}
    \label{fig:merged-mmx-87-leak-rates}
\end{figure}

Following the convention~\cite{ragab2021rage}, we provide leakage rate analysis on MMX-x87 transient execution mechanism with changing iterations and leakage granularities as shown in Figure~\ref{fig:merged-mmx-87-leak-rates}. Unlike the Masked Exceptions, we observed leakage through MMX-x87 in both \Mi we analyzed. Interestingly, in Skylake-X, the highest leakage rate, 2.3 Mb/s was achieved through 1-bit granularity and with iteration N=3. The second high was 1.9 Mb/s with 8-bit granularity. Between N=3 and N=$\sim$150, we do not observe any leakage. In Raptor Lake, the leakage appears at N>200 with up to 233 Kb/s.

\paragraph{\texttt{SERIALIZE} Instruction}
Speculative execution vulnerabilities in modern processors allow attackers to observe transient execution behavior and infer secret data. This section analyzes the behavior of the \texttt{SERIALIZE} instruction in the context of speculative execution and demonstrates how it can inadvertently facilitate information leakage under certain conditions.

Intel's documentation states that the \texttt{SERIALIZE} instruction ``serializes instruction execution, ensuring all modifications to flags, registers, and memory by previous instructions are complete before subsequent instructions are fetched and executed" \cite{intel-sdm}. This behavior includes "draining all buffered writes to memory'' \cite{intel-sdm}, providing a robust barrier for instruction execution order. However, our experiments reveal that \texttt{SERIALIZE}, when executed repeatedly alongside specific instructions, does not always halt speculative execution effectively. This observation was made during experiments designed to measure speculative execution behavior and leakage patterns.

The instruction sequences given in Listing~\ref{lst:seq_combined} were executed repeatedly, with observed effects on speculative execution:
\begin{figure}[h]
\centering
\begin{lstlisting}[caption={Sequences 1, 2, and 3: Showing various data leak behaviors through \texttt{SERIALIZE}.}, label={lst:seq_combined}]
; Sequence 1: leaks beginning of the secret
%rep N
SERIALIZE
RDGSBASE EAX
VAESDECLAST YMM0, YMM2, yword [R15]
%endrep

; Sequence 2: leaks end of the secret
%rep N
SERIALIZE
RDGSBASE EAX
%endrep

; Sequence 3: no leak
%rep N
SERIALIZE
VAESDECLAST YMM0, YMM2, yword [R15]
%endrep
\end{lstlisting}

\end{figure}

Sequence 1 leaked the beginning of the secret, revealing fragments.
Sequence 2 leaked the end of the secret, revealing fragments.
 Sequence 3 did not result in any observable data leakage.

Interestingly, the first two sequences caused a measurable increase in \texttt{RECOVERY\_CYCLES}, indicating that the Resource Allocation Table (RAT) checkpoints were recovering after speculative execution. No increase in Machine Clear counters, $\mu$code assists, or exceptions was observed.

While \texttt{SERIALIZE} is intended to act as a speculation barrier, its behavior in combination with other instructions, such as \texttt{RDGSBASE} and \texttt{VAESDECLAST}, can inadvertently permit transient execution to proceed. This transient execution creates a window where secret data can be accessed speculatively and leaked through side channels. 

\begin{figure}
    \centering
    \includegraphics[width=\linewidth]{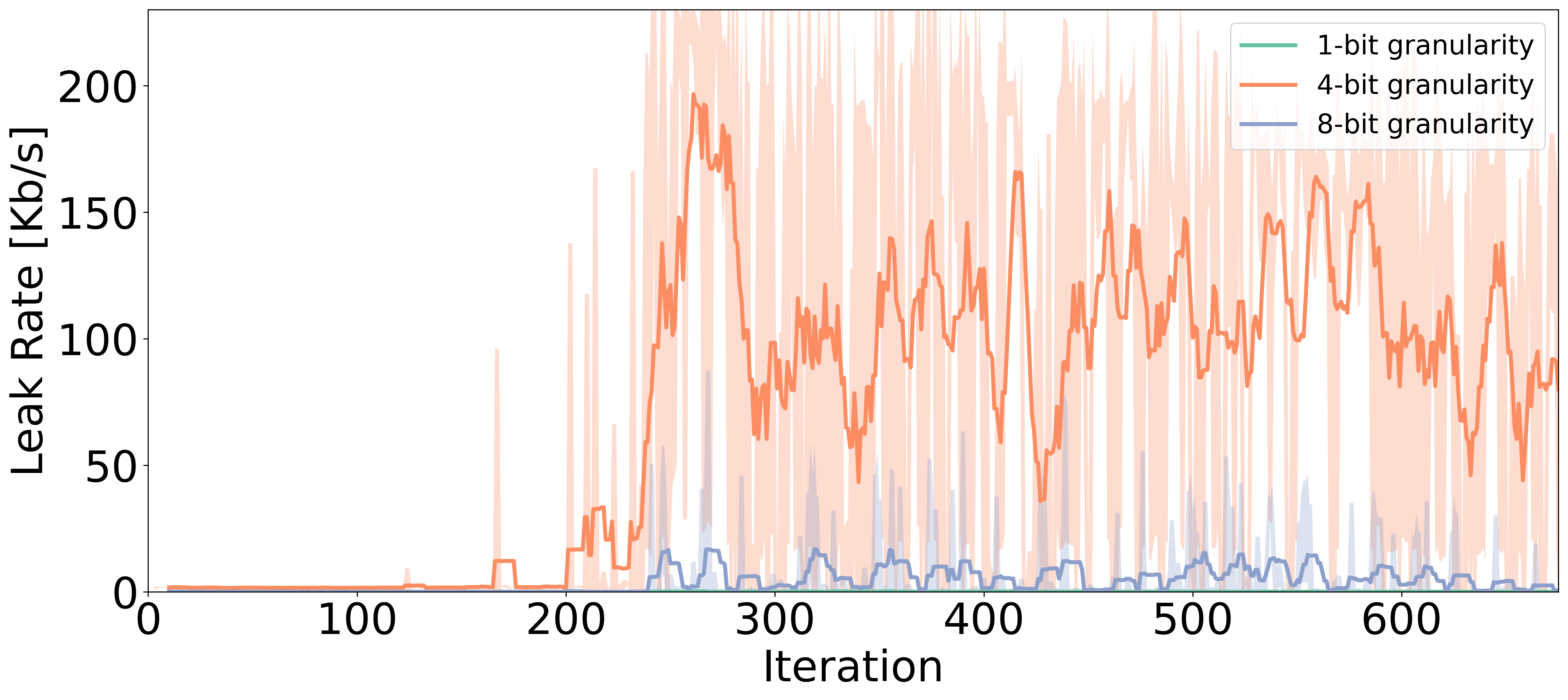}
    \caption{Leakage rate vs repetitions of the instruction pattern for \texttt{SERIALIZE} instruction measured in Raptor Lake. Darker lines are running mean and the shades are rolling variance around the mean.}
    \label{fig:serialize_leak_rate}
\end{figure}

Figure~\ref{fig:serialize_leak_rate} shows the leakage rate analysis with the changing number of iterations. The largest leakage rate observed is 230 Kb/s with 4-bit leakage granularity in Raptor Lake. We did not observe leakage in Skylake-X.

\paragraph{\texttt{VERR}/\texttt{VERW} Instructions}
The VERR (Verify Read) and VERW (Verify Write) instructions are employed to confirm if a memory segment can be read or written from the current privilege level. These instructions are crucial for security protocols, ensuring that less privileged code cannot access or alter segments belonging to more privileged levels. These instructions modify the Zero Flag based on whether a segment is readable or writable.
Although these are obsolete instructions, VERW instruction has recently given an additional functionality that wipes off the microarchitectural buffers in efforts to mitigate MDS attacks~\cite{intel2021mds} from the software.

\begin{figure}[h]
    \centering
    \begin{subfigure}[b]{0.45\textwidth}
        \centering
        \includegraphics[width=\linewidth]{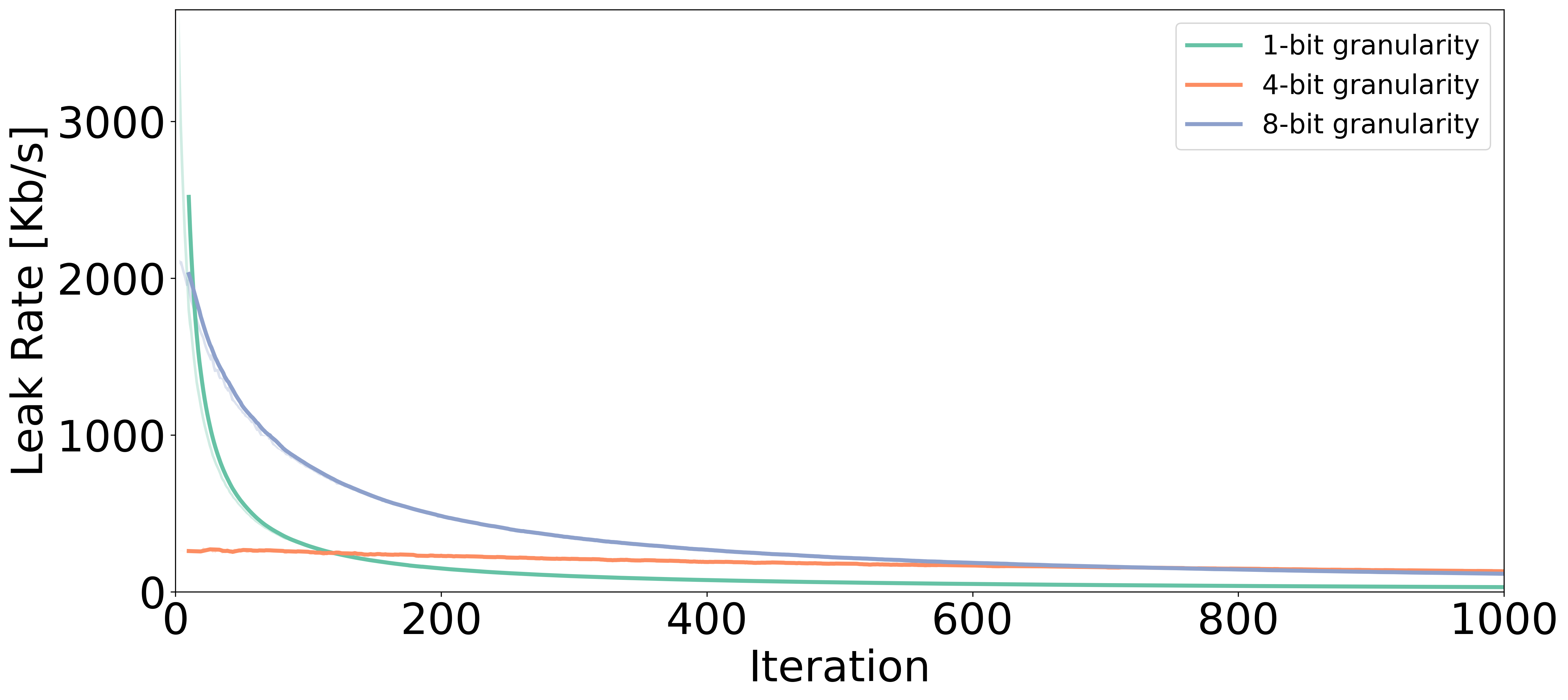}
        \caption{\texttt{VERW}}
        \label{fig:verw_list}
    \end{subfigure}
    \hfill
    \begin{subfigure}[b]{0.45\textwidth}
        \centering
        \includegraphics[width=\linewidth]{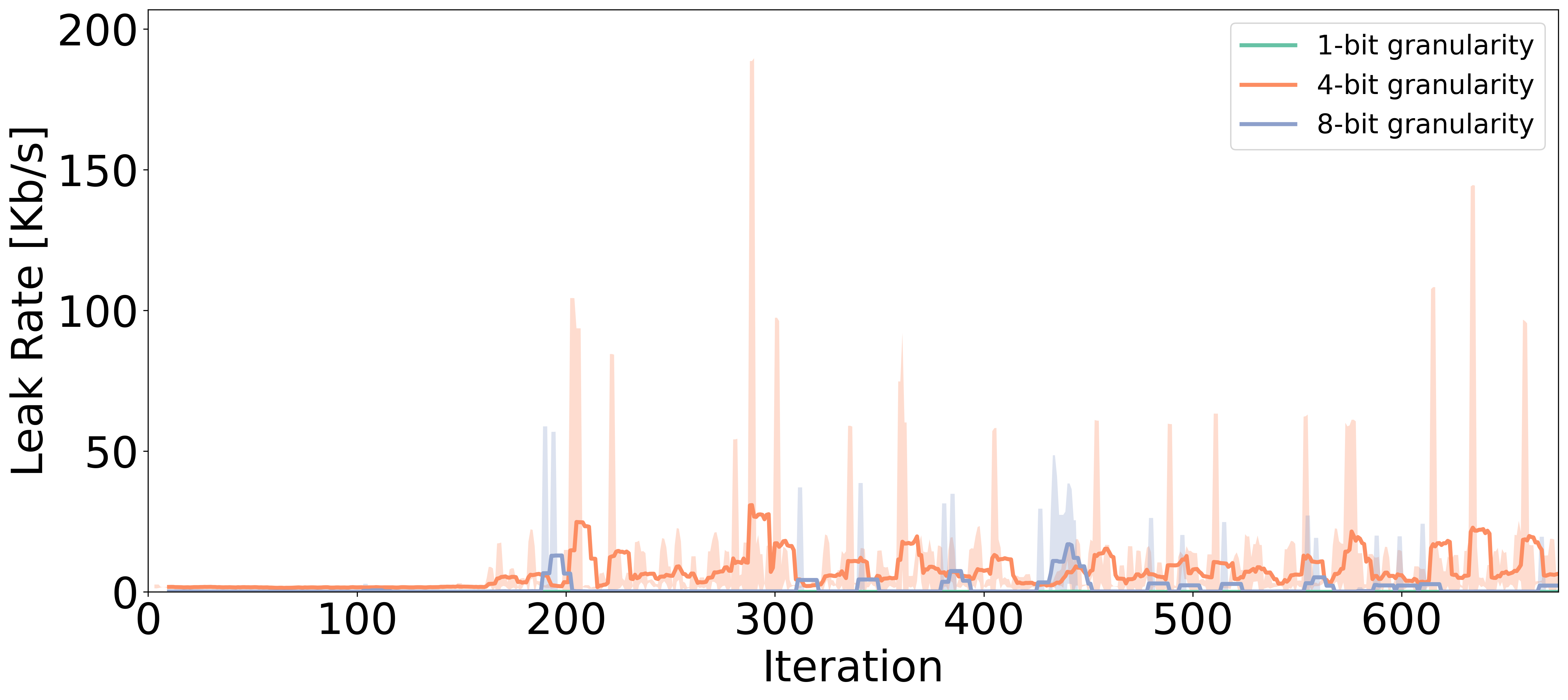}
        \caption{VERR}
        \label{fig:verr_list}
    \end{subfigure}
    \caption{Leakage rates for \texttt{VERW} and \texttt{VERR} instructions measured in Raptor Lake. In (b) Darker lines are running mean and the shades are rolling variance around the mean.}
    \label{fig:merged-leak-rates-verw-verr}
\end{figure}

$\mu$RL discovered instruction sequences given in Listing~\ref{lst:verw} that cause observable transient execution with both VERW and VERR instructions. Interestingly, we do not observe a leakage with VERW alone, without the instruction given in line 4.
On Skylake-X \Mi, VERW achieved up to 2.2 Mb/s leakage rate, and 1-bit and 8-bit leakage granularity are close to each other. The complete data is given in Appendix~\ref{sec:verw_skylake}. Note that Skylake-X is one of the microarchitectures vulnerable to MDS attacks, and a microcode patch was issued for the VERW mitigation functionality.
On Raptor Lake, VERW achieved up to 3.7 Mb/s and 2.1 Mb/s leakage rates with 1-bit and 8-bit leakage granularities, respectively. With 4-bit, we observed 279 Kb/s leakage rate.  VERR achieved up to 207 Kb/s leakage rate with 4-bit leakage granularity, only in Raptor Lake.
The relation between the iteration and leakage rates are given in Figure~\ref{fig:merged-leak-rates-verw-verr}.

\begin{figure}[h]
\begin{lstlisting}[caption=Instruction sequences with leakage through \texttt{VERW} and \texttt{VERR}., label={lst:verw}]
;leaks through VERW
%rep N
VERW word [R15]
LZCNT EDX, dword [R15]
%endrep

;leaks through VERR
%rep N
VERR AX
%endrep
\end{lstlisting}
\end{figure}



\paragraph{CLMUL Instructions}
The CLMUL instruction set is designed to accelerate cryptographic operations by performing carry-less multiplication on 128-bit operands. This is particularly useful in cryptographic algorithms like AES and GCM, where polynomial multiplications over GF($2^{128}$) are required, and carry propagation is not needed.
Listing~\ref{lst:CMUL} shows an RL-generated sequence which includes \texttt{PCLMULQDQ} which is one of the CLMUL instructions.

\begin{figure}
    \centering
    \includegraphics[width=\linewidth]{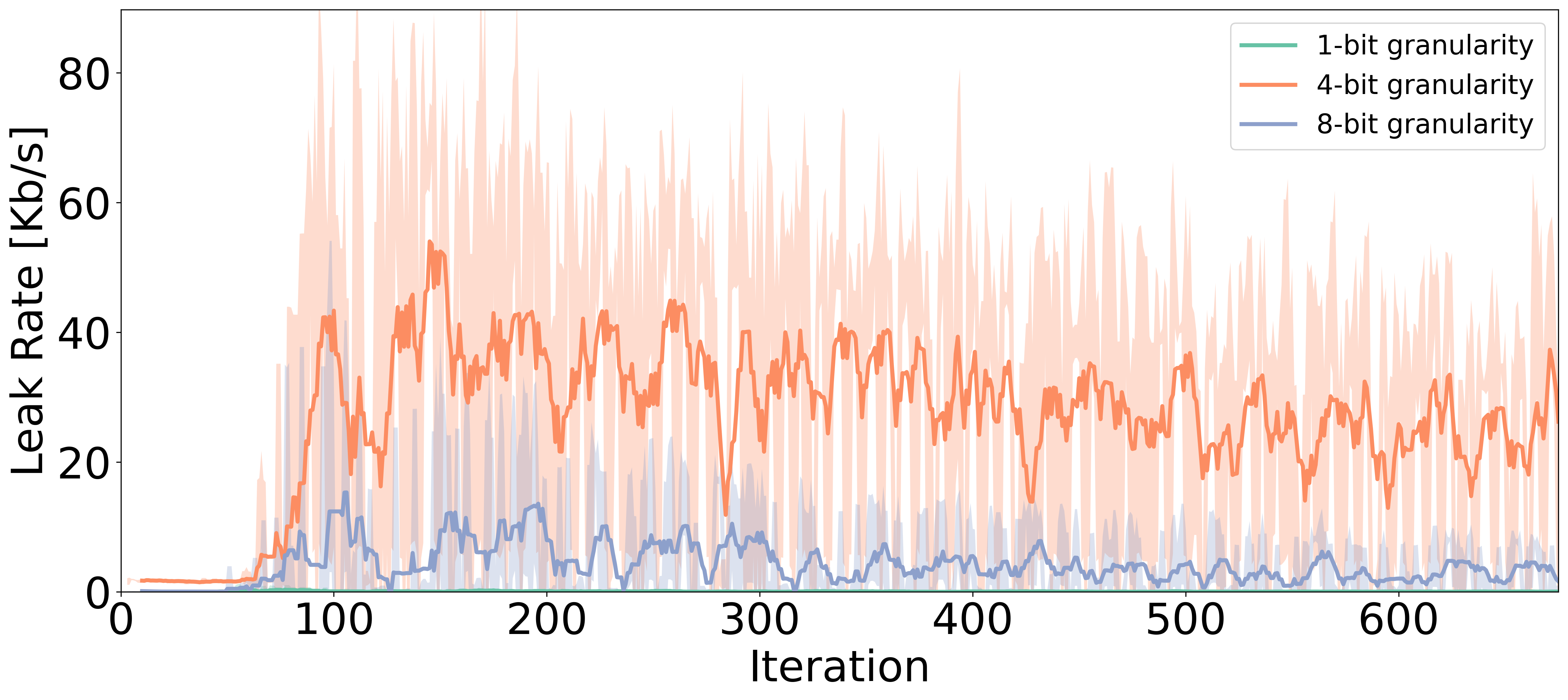}
    \caption{Leake rates for CLMUL measured in Raptor Lake. Darker lines are running mean and the shades are rolling variance around the mean.}
    \label{fig:cmul_leak_rate}
\end{figure}

We observed leakage only in Raptor Lake with up to 90 Kb/s with 4-bit granularity as shown in Figure~\ref{fig:cmul_leak_rate}.

\begin{figure}[h]
\begin{lstlisting}[caption=Instruction sequence with leakage through \texttt{CMUL}, label={lst:CMUL}]
;leaks through PCLMULQDQ
%rep N
PCLMULQDQ XMM4, [R15], 2
VCVTPS2PH XMM0, YMM0, 2
LOCK CMPXCHG16B [R15]
%endrep
\end{lstlisting}
\end{figure}

\paragraph{Miscellaneous}
We observed leakage with the combination of other instructions such as LSL+RDSCP, LAR+RDSCP, LAR+MULX, LAR+ADCX+CMOVNL, LAR+PREFETCHWT1, etc. However, to save space, we give some of them in the Appendix~\ref{sec:misc}.

\begin{table*}[t]
    \centering
    \begin{tabular}{ccccccc}
        \toprule
        Mechanism & \multicolumn{2}{c}{Skylake-X} & \multicolumn{2}{c}{Raptor Lake} \\ 
        \cmidrule(lr){2-3} \cmidrule(lr){4-5}
                  & Availability & Leakage Rate [Kb/s] & Availability & Leakage Rate [Kb/s] \\ 
        \midrule
        FP Assist~\cite{ragab2021rage} & \cmark & 222 & \cmark & 306 \\ 
        SMC~\cite{ragab2021rage} & \cmark & 235 & \cmark & 481 \\ 
        BHT~\cite{Kocher2018spectre} & \cmark & 169 & \cmark & 305 \\ 
        Exception w/ Handler~\cite{Lipp2018meltdown} & \cmark & 217 & \cmark & 280 \\ 
        TSX~\cite{Lipp2018meltdown} & \cmark & 227 & \xmark & N/A$^*$ \\ 
        MD~\cite{ragab2021rage} & \cmark & 225 & \cmark & 276 \\ 
        XMC~\cite{ragab2021rage} & \cmark & <1 & \cmark & 395 \\ 
        MO~\cite{ragab2021rage}& \cmark & 235 & \cmark & 7 \\
        Masked FP Exception \textbf{(This work)} & \cmark & 214 & \xmark & 0 \\
        MMX-x87 \textbf{(This work)} & \cmark & 213 & \cmark & 233 \\
        SERIALIZE \textbf{(This work)}& \xmark & 0 & \cmark & 230 \\ 
        VERW \textbf{(This work)}& \cmark & 218 & \cmark & 279 \\   
        VERR \textbf{(This work)}& \xmark & 0 & \cmark & 207 \\   
        CLMUL \textbf{(This work)}& \xmark & 0 & \cmark & 90 \\ 
        LSL+RDSCP \textbf{(This work)}& \xmark & 0 & \cmark & 210 \\ 
        LAR \textbf{(This work)}& \xmark & 0 & \cmark & 30 \\ 

        \bottomrule
    \end{tabular}
    \caption{Comparison of availability and leakage rates of different transient execution mechanisms in Intel Skylake-X and Raptor Lake microarchitectures with 4-bit leakage granularity.*The TSX instruction set extension is not available in Raptor Lake. The remaining \xmark ~marks show the instructions are supported, yet we do not observe leakage.}
    \label{tab:placeholder_label}
\end{table*}

\section{Exploitability of Discovered Transient Execution Mechanisms}
In this section, we show how Meltdown-like vulnerabilities can be exploited without having TSX or exception handling, thanks to discovered instruction sequences by $\mu$RL.

The original Meltdown exploit~\cite{Lipp2018meltdown}, where an attacker uses transient execution to leak data from kernel memory, has been mitigated by KPTI implemented in the Linux Kernel. However, mitigations like KPTI are application-specific, and they mitigate only one element in the attack chain, in this case, the availability of kernel addresses mapped in the virtual memory. Yet, the root cause for the transient execution has not been mitigated in the hardware. Therefore, we argue that different applications can potentially still be vulnerable and be exploited using different transient execution mechanisms. Since finding new software applications vulnerable to transient execution is not part of the scope of this work, we adopt the original Meltdown setup. 
For the proof of concept, we disable KPTI on the Linux kernel.

\begin{figure}
    \centering
    \includegraphics[width=\linewidth]{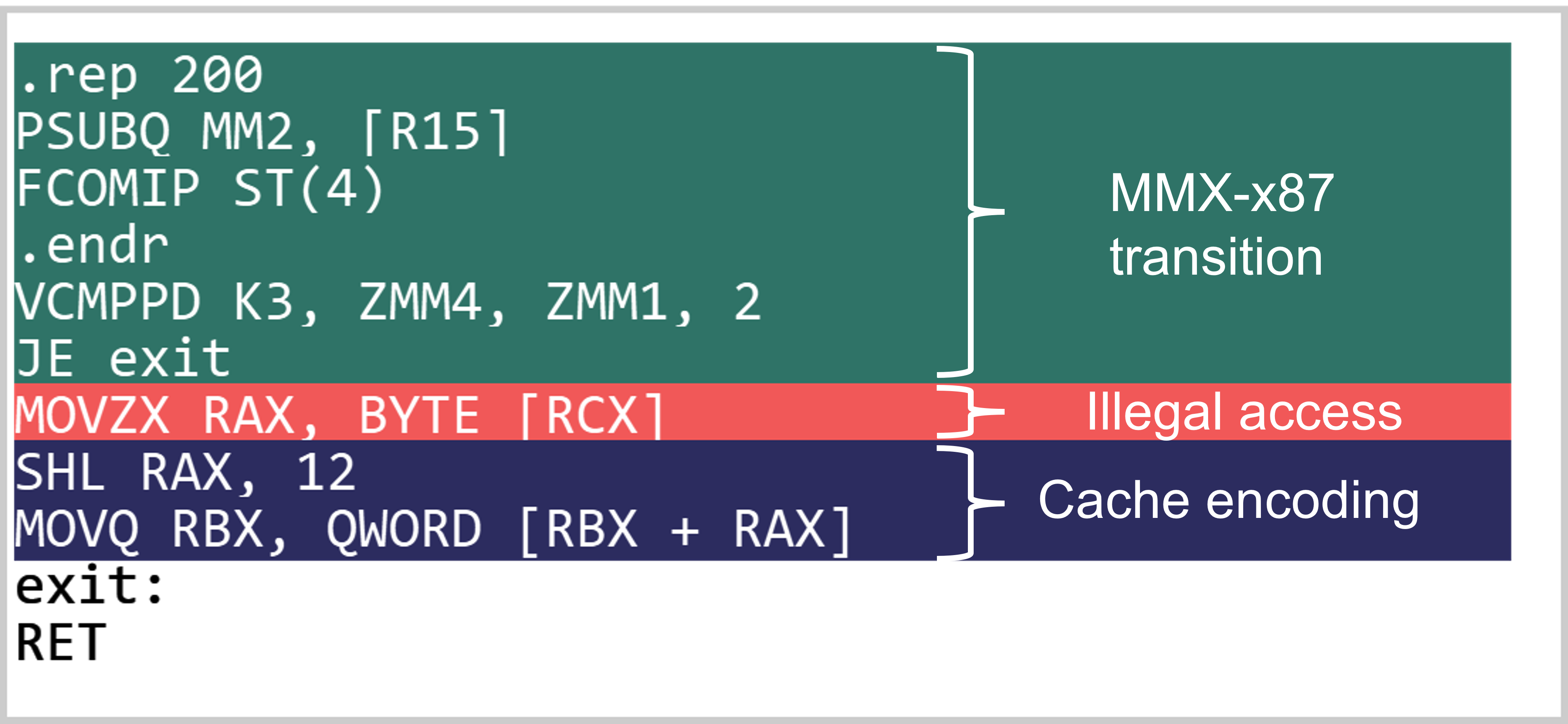}
    \caption{Proof of concept code that demonstrates the use of RL-generated MMX-x87 transient execution mechanism for reading the physical memory.}
    \label{fig:poc}
\end{figure}

Using the $\mu$RL generated instructions sequence shown in Figure~\ref{fig:poc}, we trigger transient execution of the following memory access in line, which is an illegal access to kernel memory. Speculative load from the kernel memory is then encoded into the cache so that we can decode it later using Flush+Reload.

In this experiment, we were able to read a pre-chosen secret value from the kernel memory without using a fault handler or TSX instruction set. This result demonstrates how $\mu$RL generated instructions sequences can be exploited, and they can be used as alternatives to the known attack vectors.

\section{Discussion}

\subsection{Limitations}

While our RL framework demonstrates promising results in discovering microarchitectural vulnerabilities, it is important to acknowledge its current limitations:

\paragraph{Exploration Focus}
While the framework is effective at generating instruction sequences that trigger vulnerabilities, it does not account for the entire system's attack surface. For example, vulnerabilities involving interactions between hardware and software, such as operating system mitigations or compiler optimizations, are not explicitly explored.

In this work, we did not consider the impact of other system configurations such as Hyperthreading, TSX, SGX, AVX, HW prefetch, previous mitigations, Kernel Samepage Merging, ASLR, page table layout, etc. on the \Mi vulnerabilities. We leave this for future work.

\paragraph{Sparse and Delayed Rewards}
Within the search space, only a small fraction of instruction sequences would indicate a vulnerability, assuming the design went through a thorough security review previously. The delayed nature of rewards, which depends on the cumulative effect of multiple instructions, further complicates learning. Therefore, reward signal is sparse and delayed, making it challenging for the agent to learn the optimal policy.

\paragraph{Incomplete CPU State Observability}
Our framework relies on performance counters and partial state observations to infer microarchitectural behavior. However, it cannot directly access internal processor states or transient microarchitectural effects that are not captured by these counters. This black-box approach may miss vulnerabilities that require finer-grained or privileged insights into CPU internals. More subtle vulnerabilities can potentially be detected by the deployment of $\mu$RL on processors by the chip vendors for an internal security review.
%
%
%
%
%
%
Addressing these limitations in future work will further enhance the robustness and generalizability of $\mu$RL, enabling more comprehensive microarchitectural vulnerability discovery.

\subsection{Scalability}


\paragraph{Scaling Across Vulnerabilities}
The framework can be adapted to discover diverse classes of vulnerabilities (e.g., cache-based, transient execution) by modifying the reward function to prioritize unique microarchitectural signals and by incorporating domain-specific performance counters.

\paragraph{Scaling to Different CPU Microarchitectures}
Dynamic environment generation enables the framework to model distinct architectural features of various CPUs. Transfer learning can be employed to initialize RL agents using policies trained on similar architectures, reducing the computational overhead of adapting to new designs. Indeed, we have shown that some of the vulnerabilities discovered on one CPU can be transferred to another CPU, such as MMX-x87 and \texttt{VERW}.

\paragraph{Scaling to Heterogeneous Hardware}
To support hardware like GPUs, the action space and environment must be tailored to GPU-specific instruction sets and execution models. This involves building GPU-aware environments that simulate instruction behavior and adapting performance metrics, such as memory throughput and warp divergence, as observation signals.

\paragraph{Parallel and Distributed Execution}
Scaling the framework for large-scale exploration across diverse hardware is feasible through parallelization and distributed training. By leveraging multiple nodes and cores, the framework can simultaneously evaluate instruction sequences on varied architectures.


\section{Conclusion}


In this work, we proposed an RL framework for discovering \Mi vulnerabilities, demonstrating its capability to efficiently navigate the vast and complex instruction space of modern CPUs. The framework successfully rediscovered known vulnerabilities and uncovered new transient execution mechanisms, such as masked floating-point exceptions and MMX-to-x87 transitions, showcasing its potential as a powerful tool for hardware security research.

The proposed framework is notable for its adaptability across different \Mi, leveraging custom RL environments and performance counter feedback to guide exploration. By employing hierarchical action spaces, parallelized training, and dynamic environment generation, it addresses the scalability challenges inherent in exploring diverse instruction sets and architectures. Additionally, the use of transfer learning reduces the cost of applying the framework to new processor models, making it a versatile solution for vulnerability discovery.

Beyond CPUs, the framework can be extended to analyze other hardware platforms, such as GPUs and accelerators, by adapting the action space and observation metrics to the specific execution models and vulnerabilities of those platforms. The integration of GPU-specific metrics, for instance, could open new avenues for discovering side-channel vulnerabilities in high-performance computing environments.


Overall, the proposed RL framework represents a significant step forward in automating \Mi vulnerability discovery. By scaling efficiently across vulnerabilities, architectures, and hardware platforms, it lays the groundwork for a more systematic and adaptive approach to hardware security testing, ensuring that modern processors remain resilient against emerging threats.


 




\bibliographystyle{plain}
\bibliography{references}
\appendix
\section{Explanation of CPU Performance Events}\label{sec:hpc_descriptions}
This section provides explanations for each of the CPU performance events listed in Table~\ref{tab:cpu_events}:

\begin{itemize}
    \item \textbf{UOPS\_ISSUED.ANY}: This event counts the total number of micro-operations (uops) issued by the front end of the processor. It helps to understand how often the CPU is generating work for the execution units.

    \item \textbf{UOPS\_RETIRED.RETIRE\_SLOTS}: This counter measures the number of micro-operations that have been retired, representing the slots used in the retirement stage. High values indicate effective utilization of the CPU pipeline.

    \item \textbf{INT\_MISC.RECOVERY\_CYCLES\_ANY}: This event counts the cycles the processor spends in recovery due to issues in the integer pipeline, such as branch mispredictions. It provides insight into potential inefficiencies in the execution flow.

    \item \textbf{MACHINE\_CLEARS.COUNT}: This counter tracks the total number of machine clears. Machine clears occur when the processor needs to flush the pipeline, often due to errors or interruptions, impacting overall performance.

    \item \textbf{MACHINE\_CLEARS.SMC}: This event counts machine clears specifically triggered by self-modifying code. Self-modifying code requires the processor to invalidate instructions and restart, which is costly in terms of performance.

    \item \textbf{MACHINE\_CLEARS.MEMORY\_ORDERING}: This counter registers machine clears due to memory ordering conflicts. Such conflicts require the CPU to reset and reorder memory accesses, which can degrade performance in multithreaded applications.

    \item \textbf{FP\_ASSIST.ANY}: This event counts floating-point assists, which are special handling operations needed to process floating-point instructions. High counts may indicate heavy floating-point computation workloads or suboptimal code for floating-point operations.

    \item \textbf{OTHER\_ASSISTS.ANY}: This counter registers other assist events, including exceptions and corrections for specific instructions or situations. This can give insight into issues like misaligned memory access or uncommon instruction usage.

    \item \textbf{CPU\_CLK\_UNHALTED.ONE\_THREAD\_ACTIVE}: This event measures the number of cycles during which at least one thread is active. This counter is useful for understanding overall CPU utilization, especially in multi-threaded environments.

    \item \textbf{CPU\_CLK\_UNHALTED.THREAD}: This counter measures the cycles during which a specific thread remains active. It provides data on individual thread activity and allows for a more granular view of CPU utilization.

    \item \textbf{HLE\_RETIRED.ABORTED\_UNFRIENDLY}: This event counts the hardware lock elision (HLE) transactions that were aborted due to “unfriendly” reasons, such as interference by other threads or incompatible instructions. High values can indicate issues with lock-based concurrency.

    \item \textbf{HW\_INTERRUPTS.RECEIVED}: This counter measures the number of hardware interrupts received by the processor. Hardware interrupts are signals from external devices that require immediate attention, potentially affecting processor performance by disrupting normal execution flow.
\end{itemize}
\section{Leakage Rates Analysis on Skylake}\label{sec:verw_skylake}

\begin{figure}
    \centering
    \includegraphics[width=\linewidth]{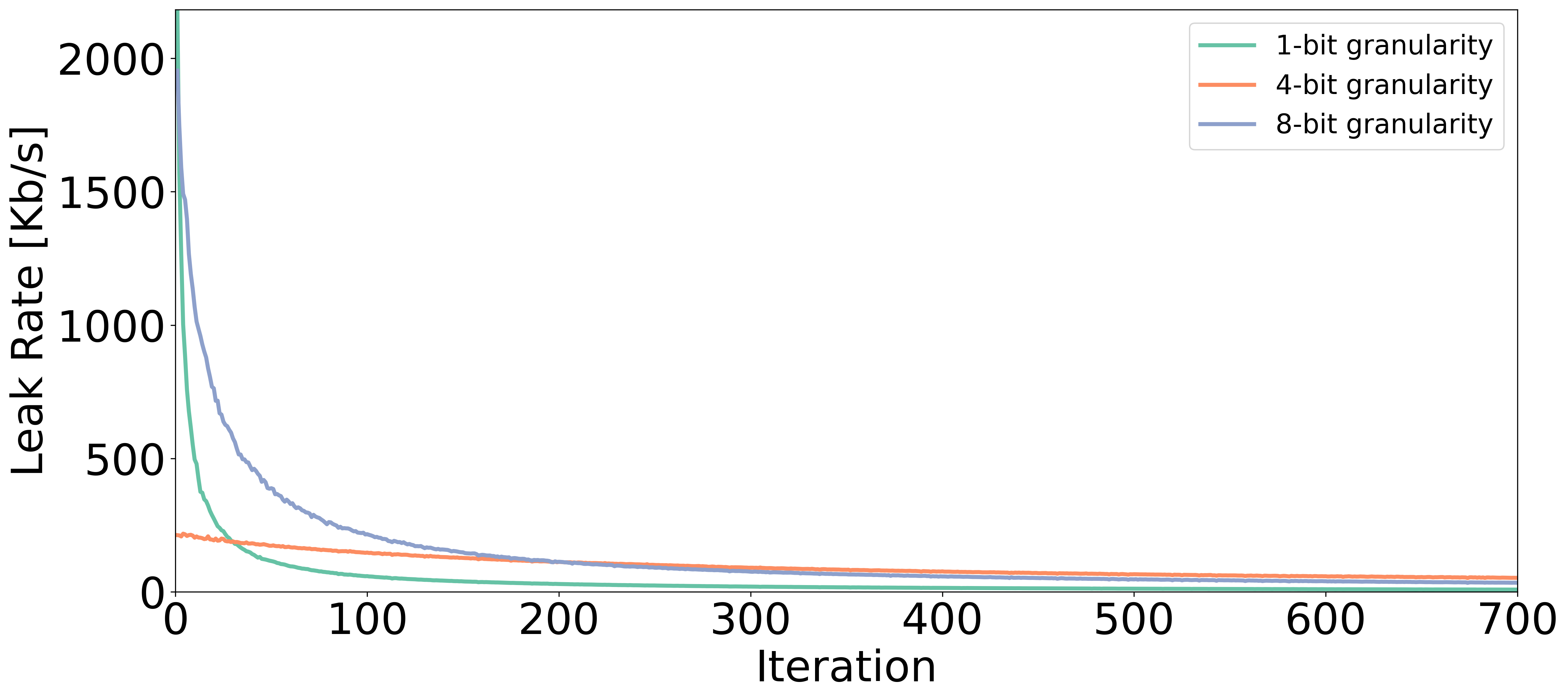}
    \caption{Leak rates for VERW measured in Skylake-X.}
    \label{fig:verwwww}
\end{figure}

Leak rates for VERW measured in Skylake-X is given in Figure~\ref{fig:verwwww}.

\section{Other $\mu$RL-discovered Transient Execution Mechanisms}\label{sec:misc}

\begin{figure}
    \centering
    \includegraphics[width=\linewidth]{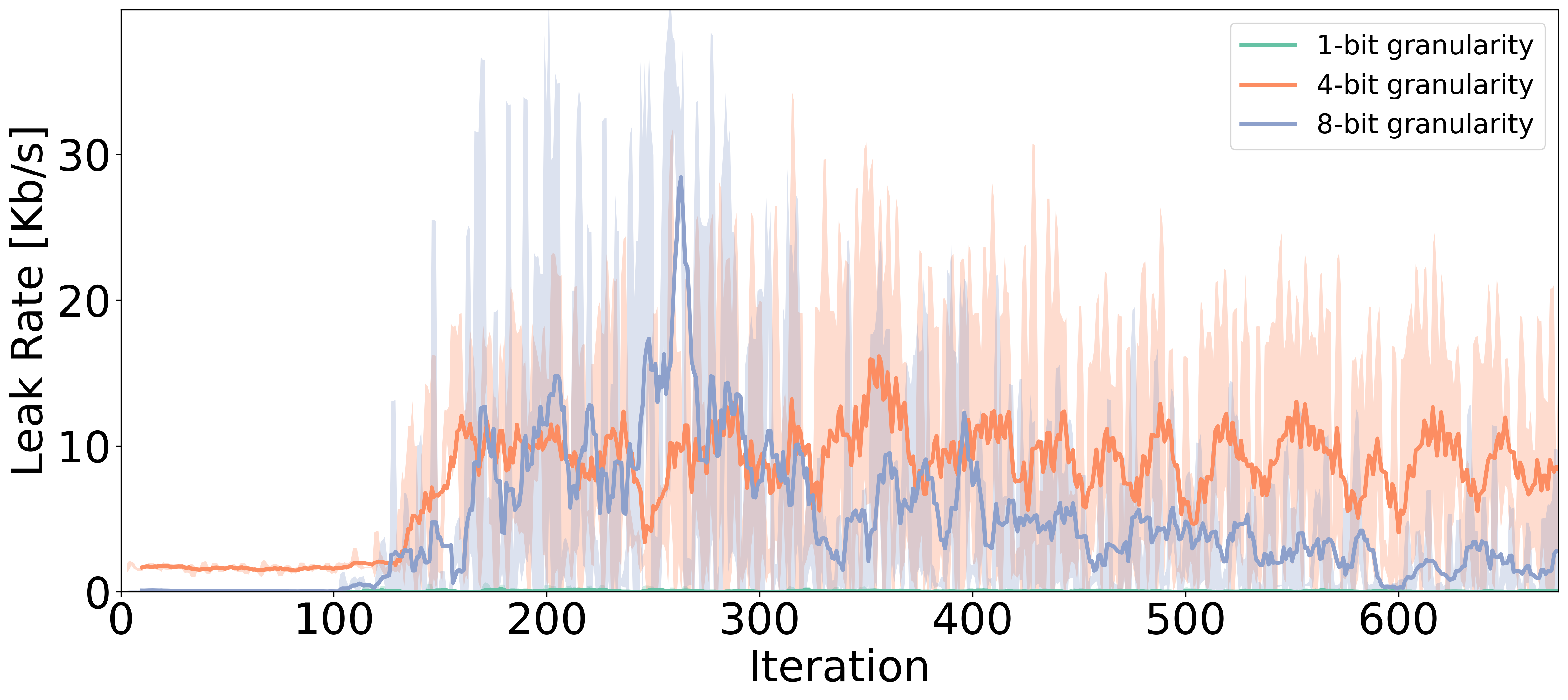}
    \caption{Leak rates for LAR measured in Raptor Lake.}
    \label{fig:larrrrr}
\end{figure}

\paragraph{LAR}
The LAR instruction loads the access rights of a segment into a register, based on the segment selector provided. It is useful for managing memory segmentation and ensuring the correct privilege levels when accessing different memory regions.
Leak rates for LAR measured in Raptor Lake in Figure~\ref{fig:larrrrr}. Four instruction sequences with \texttt{LAR} leakage are given in Listing~\ref{lst:lar}.

\begin{figure}[h]
\begin{lstlisting}[caption=Four instruction sequences with \texttt{LAR} leakage, label={lst:lar}]
; seq 1: leaks through LAR
%rep 500
LAR RAX, [R15]
MULX RAX, RAX, qword [R15]
%endrep
; seq 2: leaks through LAR
%rep 500
LAR RCX, [R15]
ADCX RCX, qword [R15]
CMOVNL EAX, EDX
%endrep
; seq 3: leaks through LAR
%rep 500
LAR AX, [R15]
RDTSCP
%endrep
; seq 4: more stable leakage with prefetchwt1
%rep 500
PREFETCHWT1 byte [R15]
LAR ESP, [R15]
LZCNT EBX, dword [R15]
%endrep
; seq 5: no dependency
%rep 500
LAR DX, DX
LOCK CMPXCHG16B [R15]
%endrep
; seq 6: no dependency
%rep 500
WRGSBASE RSP
CMPXCHG16B [R15]
LAR DX, AX
TZCNT RDX, qword [R15]
PEXT RBX, RDX, RAX
%endrep
\end{lstlisting}
\end{figure}







\begin{figure}[h]
    \centering
    \includegraphics[width=\linewidth]{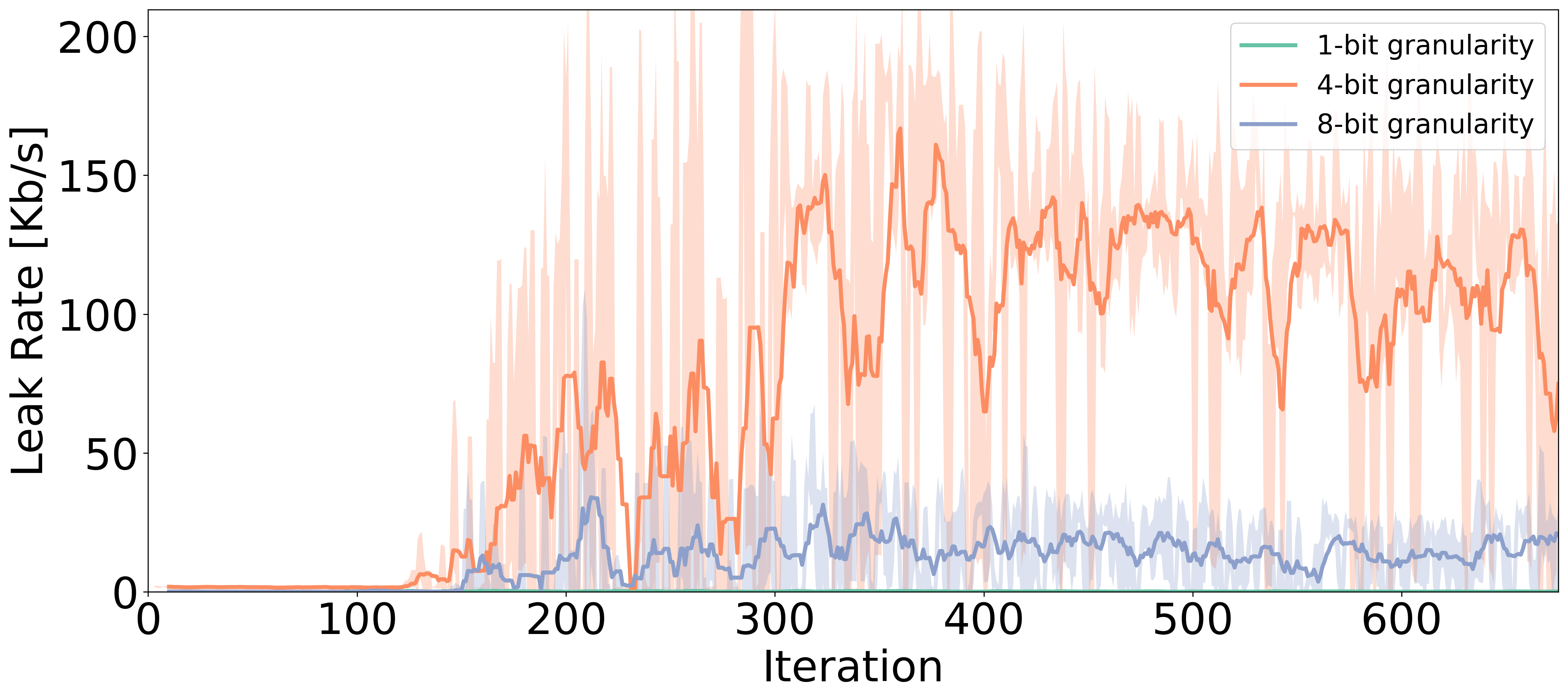}
    \caption{Leak rates for LSL+RDTSCP measured in Raptor Lake.}
    \label{fig:lslllllll}
\end{figure}

\paragraph{LSL+RDSP}
The LSL instruction loads the memory segment limit from a specified segment selector into a register, which is useful for segmentation tasks by providing the size of a memory segment, helping with boundary checks. Leak rates for LSL+RDTSCP measured in Raptor Lake.
Leak rates for LSL+RDTSCP measured in Raptor Lake are given in Figure~\ref{fig:lslllllll}.

\section{Leakage Rate Analysis on Previously Known Transient Execution Mechanisms}
\begin{figure}[h]
    \centering
    \includegraphics[width=\linewidth]{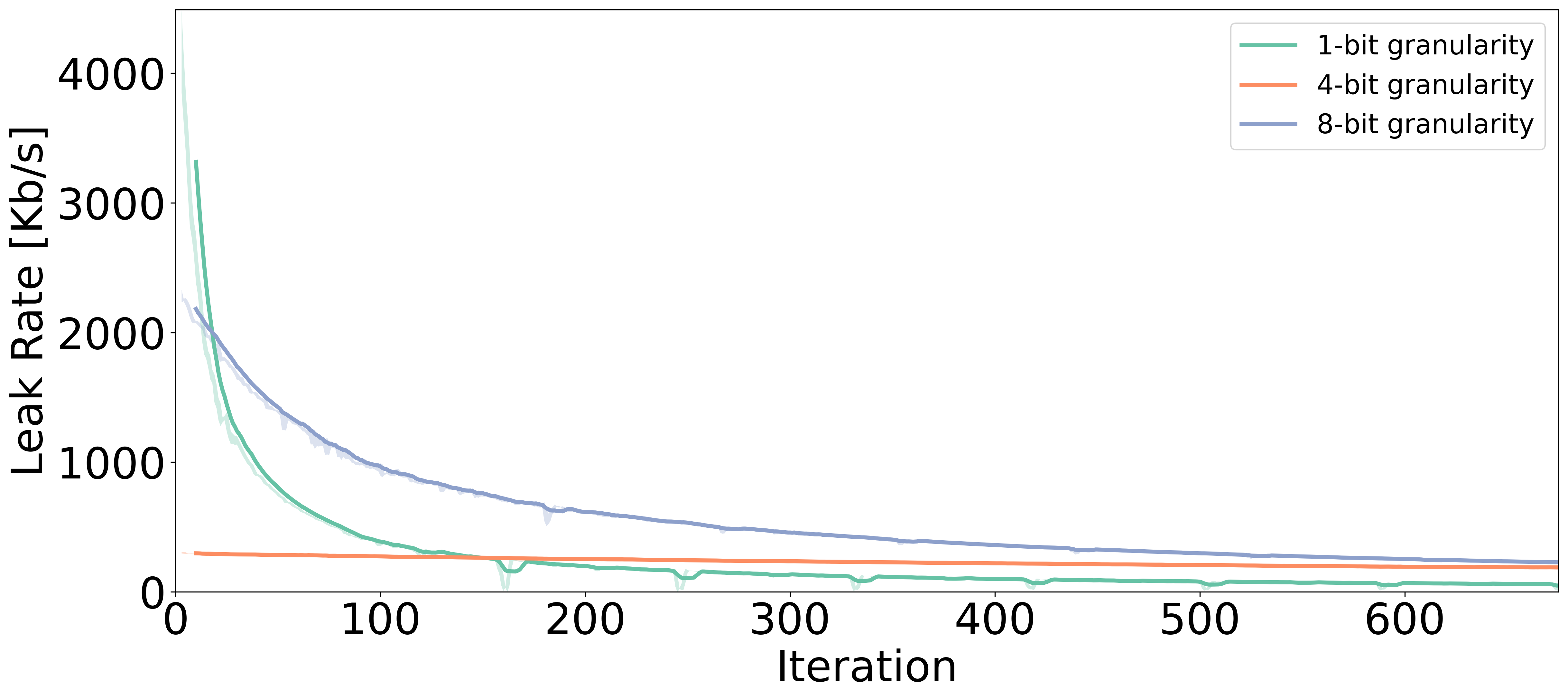}
    \caption{Leak rates for FP assist measured in Raptor Lake.}
    \label{fig:11}
\end{figure}

\begin{figure}[h]
    \centering
    \includegraphics[width=\linewidth]{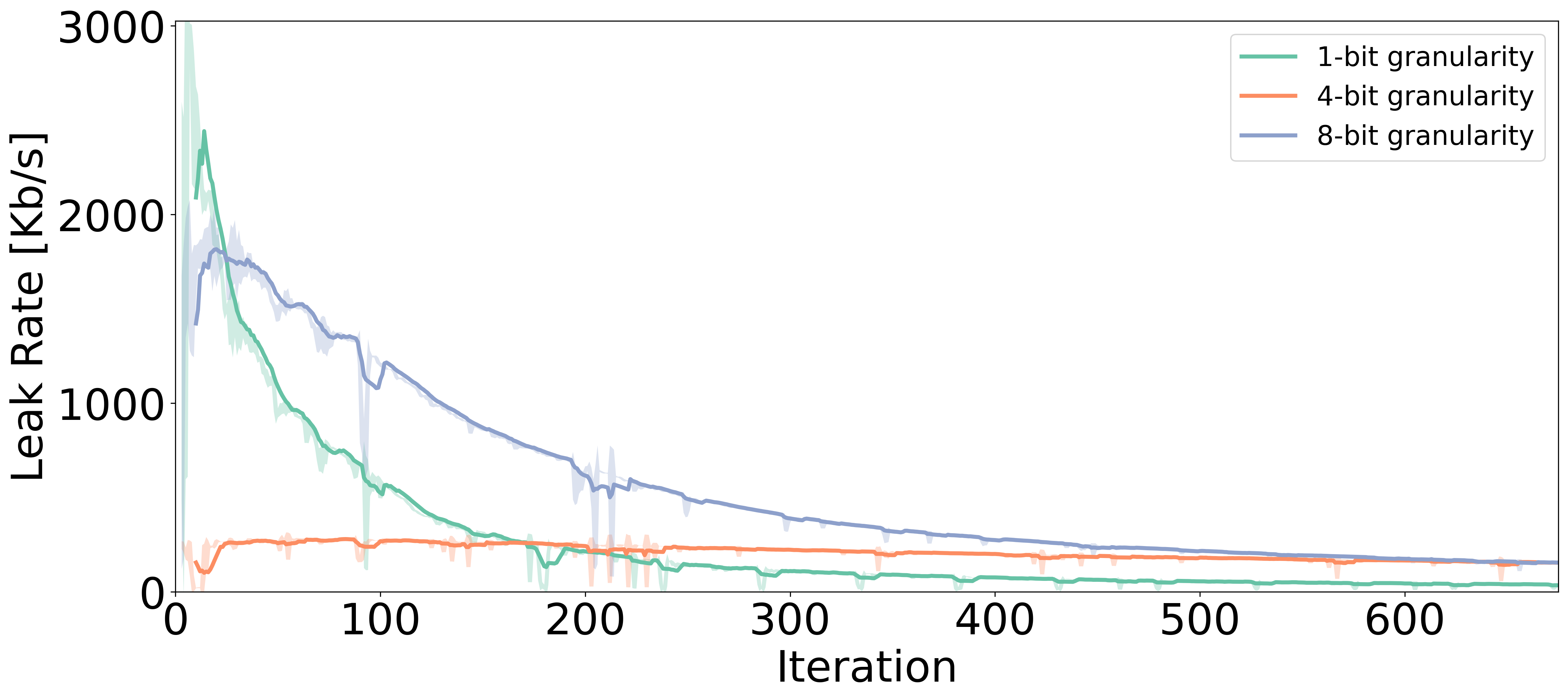}
    \caption{Leak rates for Memory disambiguation measured in Raptor Lake.}
    \label{fig:12}
\end{figure}

\begin{figure}[h]
    \centering
    \includegraphics[width=\linewidth]{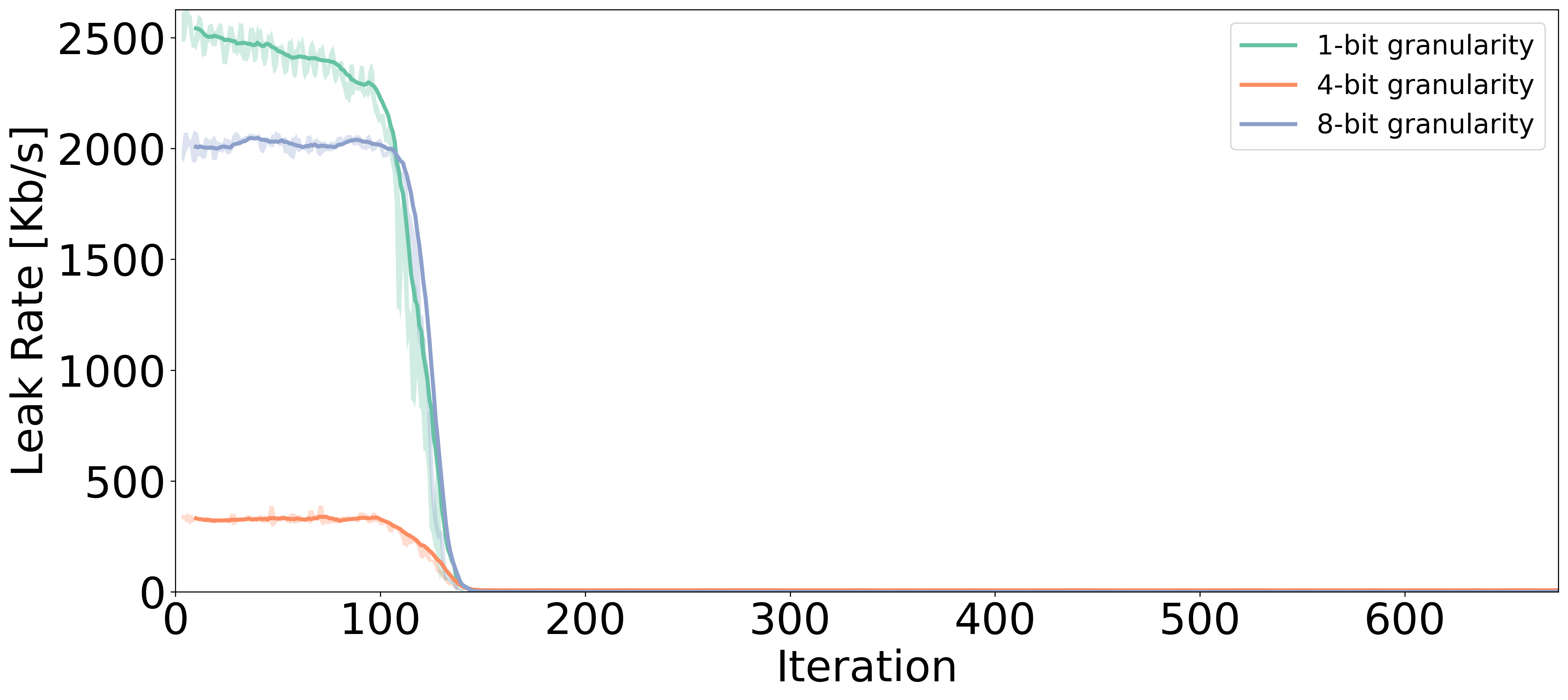}
    \caption{Leak rates for Memory ordering measured in Raptor Lake.}
    \label{fig:13}
\end{figure}

\begin{figure}[h]
    \centering
    \includegraphics[width=\linewidth]{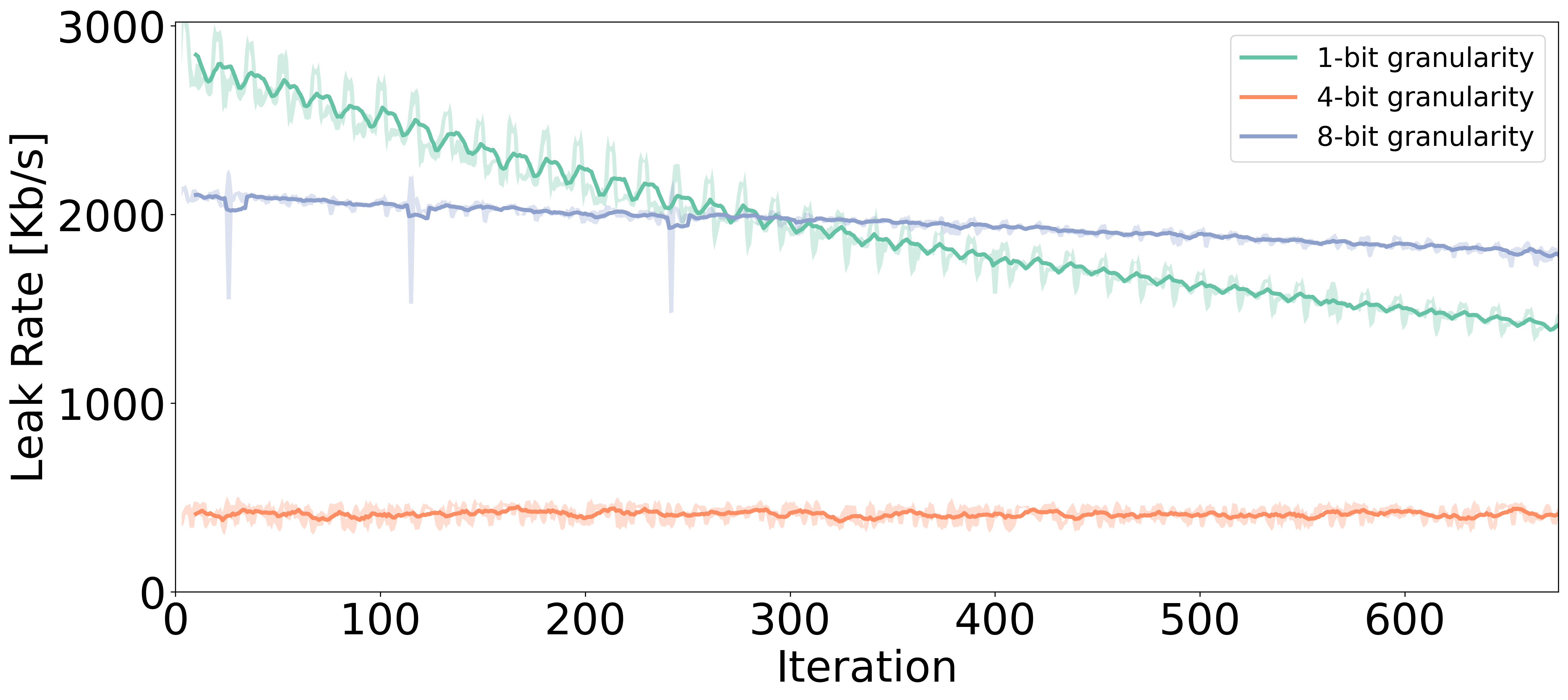}
    \caption{Leak rates for SMC measured in Raptor Lake.}
    \label{fig:14}
\end{figure}

Leakage rates for previously known mechanisms such as FP assist, MD, MO and SMC on Raptor lake are given in Figure~\ref{fig:11},~\ref{fig:12},~\ref{fig:13} and ~\ref{fig:14}.

\section{Instruction Sets}

\begin{table}[ht]
    \centering
    \footnotesize
    \begin{tabular}{|l|r|l|r|}
    \hline
    \textbf{Instruction Set} & \textbf{Count} & \textbf{Instruction Set} & \textbf{Count} \\
    \hline
    ADOX\_ADCX & 8 & AES & 12 \\
    AVX & 695 & AVX2 & 286 \\
    AVX2GATHER & 16 & AVX512F\_512 & 2192 \\
    AVX512F\_128 & 1816 & AVX512F\_256 & 1940 \\
    AVX512F\_SCALAR & 584 & AVX512DQ\_128 & 247 \\
    AVX512DQ\_256 & 281 & AVX512DQ\_512 & 357 \\
    AVX512BW\_128 & 467 & AVX512BW\_256 & 467 \\
    AVX512BW\_512 & 467 & AVX512F\_128N & 23 \\
    AVX512DQ\_SCALAR & 44 & AVX512CD\_512 & 38 \\
    AVX512CD\_128 & 38 & AVX512CD\_256 & 38 \\
    AVX512BW\_128N & 8 & AVX512DQ\_128N & 8 \\
    AVX512DQ\_KOP & 18 & AVX512BW\_KOP & 34 \\
    AVX512F\_KOP & 15 & AVXAES & 12 \\
    I86 & 809 & I386 & 196 \\
    I486REAL & 37 & CMOV & 96 \\
    PENTIUMREAL & 5 & I186 & 124 \\
    LONGMODE & 24 & LAHF & 2 \\
    I286PROTECTED & 26 & I286REAL & 10 \\
    FAT\_NOP & 3 & RDPMC & 1 \\
    PPRO & 2 & BMI1 & 26 \\
    BMI2 & 32 & CET & 2 \\
    F16C & 8 & FMA & 192 \\
    INVPCID & 1 & CMPXCHG16B & 2 \\
    LZCNT & 6 & PENTIUMMMX & 129 \\
    SSE & 97 & MOVBE & 6 \\
    PCLMULQDQ & 2 & RDRAND & 3 \\
    RDSEED & 3 & RDTSCP & 1 \\
    RDWRFSGS & 8 & FXSAVE & 2 \\
    FXSAVE64 & 2 & SSEMXCSR & 2 \\
    SSE2 & 264 & SSE2MMX & 6 \\
    SSE3 & 20 & SSE3X87 & 2 \\
    SSE4 & 96 & SSE42 & 25 \\
    POPCNT & 6 & SSSE3MMX & 32 \\
    SSSE3 & 32 & X87 & 119 \\
    FCMOV & 8 & FCOMI & 4 \\
    XSAVE & 6 & XSAVEC & 2 \\
    XSAVEOPT & 2 & XSAVES & 4 \\
    \hline
    \end{tabular}\label{tab:instruction_sets}
    \caption{Number of instructions per set used in the action space for Skylake-X.}
    \end{table}

\end{document}